\pdfoutput=1
\documentclass[11pt,a4paper]{article}
\usepackage{jheppub}
\usepackage[caption=false]{subfig}
\usepackage{graphicx}
\usepackage{hyperref}
\usepackage{multirow}
\usepackage{color}
\usepackage{amsthm}

\def\p{\partial}

\def\=:{=\hspace{-.7em}\raisebox{1.1ex}{.}\hspace{.1em}\raisebox{-0.2ex}{.}}

\newcommand{\beq}{\begin{eqnarray}}
\newcommand{\eeq}{\end{eqnarray}}
\newcommand{\non}{\nonumber\\}

\newcommand{\tr}{{\rm tr}\,}
\newcommand{\bphi}{\boldsymbol{\phi}}
\newcommand{\bpsi}{\boldsymbol{\psi}}
\newcommand{\bx}{\mathbf{x}}
\newcommand{\ol}[1]{\mkern2.75mu\overline{\mkern-2.75mu#1\mkern-1.5mu}\mkern1.5mu}
\renewcommand{\d}{{\mathrm{d}}}
\renewcommand{\i}{\mathrm{i}}

\newtheorem{theorem}{Theorem}
\newtheorem{corollary}{Corollary}
\newtheorem{conjecture}{Conjecture}

\begin{document}

\title{Linking number of vortices as baryon number}

\author{Sven Bjarke Gudnason$^1$ and}
\affiliation{$^1$Institute of Contemporary Mathematics, School of
  Mathematics and Statistics, Henan University, Kaifeng, Henan 475004,
  P.~R.~China} 
\author{Muneto Nitta$^2$}
\affiliation{$^2$Department of Physics, and Research and Education
  Center for Natural Sciences, Keio University, Hiyoshi 4-1-1,
  Yokohama, Kanagawa 223-8521, Japan}
\emailAdd{gudnason(at)henu.edu.cn}
\emailAdd{nitta(at)phys-h.keio.ac.jp}

\abstract{
We show that the topological degree of a Skyrmion field is the same as
the Hopf charge of the field under the Hopf map and thus equals the
linking number of the preimages of two points on the 2-sphere under
the Hopf map. We further interpret two particular points on the
2-sphere as vortex zeros and the linking of these zero lines follows
from the latter theorem.
Finally we conjecture that the topological degree of the Skyrmion can
be interpreted as the product of winding numbers of vortices
corresponding to the zero lines, summing over clusters of vortices. 
}


\maketitle

\section{Introduction}

Skyrmions are topological solitons \cite{Manton:2004} of the texture
type, i.e.~they are maps from one-point compactified 3-space,
$X=\mathbb{R}^3\cup\{\infty\}\simeq S^3$ to a target space $N=S^3$
with a nonvanishing topological degree
$\pi_3(S^3)=\mathbb{Z}\ni B\neq 0$. 
Usually the map is constructed using an SU(2) matrix $U$, where the
nonlinear sigma model constraint is $\det U=1$, which forces the 4
components to live on a 3-sphere of unit radius.
It is also possible to write the SU(2) matrix as an O(4) vector of
unit length.
In this paper, however, it proves convenient to write the SU(2) field
as two complex scalar fields, $\psi_{1,2}$,  
living on the complexified 1-sphere ($|\psi_1|^2+|\psi_2|^2=1$).
The convenience is two-fold.
First of all, we would like to associate the zero lines of each
complex scalar field with (deformed) vortex rings.
Secondly, it proves convenient for our calculations as we will be
using the Hopf map, which is naturally written in terms of two complex 
scalar fields. 

First we prove a theorem which shows that under the Hopf map, the
map $\bpsi:\mathbb{R}^3\cup\{\infty\}\simeq S^3\to S^3$ of degree $B$
will necessarily have Hopf charge $Q=B$. 
This statement is known in the
literature \cite{Ward:2001vi,Manton:2004} and has been used several
times to generate initial conditions for
Hopfions \cite{Battye:1998pe,Battye:1998zn} 
in a different model, called the Faddeev-Skyrme
model \cite{Faddeev:1996zj}, which maps $\mathbb{R}^3$ to $S^2$ and
thus naturally possesses a Hopf charge. 
Nevertheless, we have not found the theorem written down in the
literature, and thus we shall give it here and supply a
proof.\footnote{Ref.~\cite{Meissner:1985nb} restricts $\tr U=0$, and
thus maps $\mathbb{R}^3\to S^2$ and not $\mathbb{R}^3\to S^3$;
therefore we do not consider the calculation of the Hopf charge there
as a general proof.
Similarly, ref.~\cite{Ward:2004gr} finds an interpolation between the
Skyrme model and the Faddeev-Skyrme model and states that the baryon
charge equals the Hopf charge when the model is restricted to the
Faddeev-Skyrme model, i.e.~$\mathbb{R}^3\to S^2$.
We do not make such restriction in this paper. }

The implication of the theorem is that 2 distinct regular points under
the projection of a Skyrme map to the 2-sphere have preimages in
3-space with linking number $Q=B$.
We further make the interpretation of two antipodal points on the
2-sphere being vortex zeros. So far all is done with rigor.
Finally, we conjecture that we can interpret the topological degree of
a Skyrmion map as the product of winding numbers of two vortex lines,
summing over clusters of wound vortices.

This paper is organized as follows.
In sec.~\ref{sec:maps}, after giving the maps, we present our theorem, 
corollary and conjecture.
In sec.~\ref{sec:examples}, we illustrate the theorem and conjecture
with examples of a toroidal vortex and rational map Skyrmions. 
Sec.~\ref{sec:discussion} is devoted to discussion and outlook.

\section{The maps}\label{sec:maps}

\subsection{Theorem and conjecture}

We begin with considering a map from $U: X\to N$ where
$X=\mathbb{R}^3\cup\,\{\infty\}\simeq S^3$ the one-point compactified
3-dimensional configuration space and $N=S^3$ is the target space,
which we take to be the 3-sphere in this paper.
Each space has an associated metric, that is $(X,g)$ and $(N,h)$. 
The map $U$ is characterized by the third homotopy group,
$B\in\pi_3(S^3)=\mathbb{Z}$ with $B$ the topological degree, which is
usually called the baryon number.

Next, we will consider the Hopf map $H: S^3\to S^2$, which is due to
the Hopf fibration $S^1\hookrightarrow S^3\overset{H}{\to}S^2$. 
The explicit form of the Hopf map is
\beq
H^a(\bpsi,\ol{\bpsi}) = \bpsi^\dag \tau^a \bpsi, \qquad
a=1,2,3, \label{eq:Hopfmap}
\eeq
with $\bpsi$ living on the complexified 1-sphere: 
\begin{align}
\bpsi &=
\begin{pmatrix}
\psi_1\\
\psi_2\\
\end{pmatrix}, \qquad
\psi_{1,2}\in\mathbb{C}, \label{eq:bpsi_def}\\
&\bpsi^\dag\bpsi = |\psi_1|^2 + |\psi_2|^2 = 1, \label{eq:S3constraint}
\end{align}
which is exactly a real 3-sphere
and $\tau^a$ are the Pauli SU(2) matrices.
The topological charge of the Hopf map is
\beq
Q \in \pi_3(S^2),
\eeq
but it is not the degree of the mapping as it is a mapping between
spaces of different dimensions.

The map $U:X\to N=S^3$ is given by
\beq
U(\mathbf{x}) =
\begin{pmatrix}
\bpsi & -\i\tau^2\ol{\bpsi}
\end{pmatrix}
=
\begin{pmatrix}
\psi_1 & -\ol{\psi}_2\\
\psi_2 & \ol{\psi}_1
\end{pmatrix}, \label{eq:Udef}
\eeq
which thus maps $\mathbb{R}^3\cup\{\infty\}=X\to S^3$, due to the
constraint \eqref{eq:S3constraint}.
The degree of the mapping $U$ from $X$ to $N$ can be calculated as the
pullback of the normalized volume form on $N$, $\Omega_N$ 
by $U$: 
\begin{align}
B &= \int_X U^*\Omega_N \non
  &= -\frac{1}{24\pi^2}\int_X 
  \tr\left(U^\dag\p_i U U^\dag\p_j U U^\dag\p_k U\right)
  \d{x}^i\wedge\d{x}^j\wedge\d{x}^k\non
  &= \frac{1}{4\pi^2}\int_X
  (\bpsi^\dag\p_i\bpsi)(\p_j\bpsi^\dag\p_k\bpsi)\;
  \d{x}^i\wedge\d{x}^j\wedge\d{x}^k. \label{eq:B}
\end{align}

Finally, we are interested in the map
$\bphi\equiv H\circ\, U: X\to S^2$, which is the composite map of $U$ 
and $H$.
This takes a field configuration on $X$, maps it with degree $B$ to
$N$ and then to $S^2$.

The Hopf charge (or Hopf index) of the above described map, $\bphi$,
is given by \cite{Gladikowski:1996mb,Faddeev:1996zj}, 
\beq
Q = -\frac{1}{4\pi^2}\int_X A\wedge F,
\label{eq:Qcharge}
\eeq
where the field-strength tensor in terms of the coordinates on $S^2$
is \cite{Gladikowski:1996mb,Faddeev:1996zj},
\beq
F = \frac14\bphi\cdot\p_i\bphi\times\p_j\bphi\;
  \d{x}^i\wedge\d{x}^j, \label{eq:FijS2}
\eeq
and $A$ is a corresponding gauge field $F=\d{A}$.
However, it is not possible to write a \emph{local} expression for the  
Chern-Simon action \eqref{eq:Qcharge} in terms of the coordinates,
$\bphi$, on $S^2$, because it vanishes identically, as well known. 

\begin{theorem}\label{thm:1}
A map $U:\mathbb{R}^3\cup\{\infty\}\to S^3$ with topological degree
$B$ under the Hopf map $H:S^3\to S^2$ has Hopf charge $Q=B$ and thus
distinct regular points on $S^2$ under the composite map
$H\circ U:\mathbb{R}^3\cup\{\infty\}\to S^2$ have preimages on
$\mathbb{R}^3\cup\{\infty\}$ that are linked $Q=B$ times. 
\end{theorem}

\noindent\emph{Proof}:
We calculate the field-strength tensor \eqref{eq:FijS2} in terms of
the coordinates on $S^3$ via the Hopf map \eqref{eq:Hopfmap} as
\begin{align}
F &= \frac{1}{4}\epsilon^{a b c}
  (\bpsi^\dag\tau^a\bpsi)\p_i(\bpsi^\dag\tau^b\bpsi)\p_j(\bpsi^\dag\tau^c\bpsi)\;
  \d{x}^i\wedge\d{x}^j\non
  &= \i\left(
    \psi_2\ol{\psi}_2\p_{i}\psi_1\p_{j}\ol{\psi}_1
    -\psi_2\ol{\psi}_1\p_{i}\psi_1\p_{j}\ol{\psi}_2
    +\psi_1\ol{\psi}_1\p_{i}\psi_2\p_{j}\ol{\psi}_2
    -\psi_1\ol{\psi}_2\p_{i}\psi_2\p_{j}\ol{\psi}_1
    \right) \d{x}^i\wedge\d{x}^j \non
  &= -\i\p_{i}\bpsi^\dag\p_{j}\bpsi\;
    \d{x}^i\wedge\d{x}^j, \label{eq:FijS3}
\end{align}
where we have used the constraint \eqref{eq:S3constraint}. 
The above-calculated field-strength tensor can also readily be
obtained from the following gauge field
\beq
A = -\frac{\i}{2}\left(\bpsi^\dag\p_i\bpsi
  - \p_i\bpsi^\dag\bpsi\right) \d{x}^i. \label{eq:AiS3}
\eeq
We can now explicitly evaluate the Hopf charge \eqref{eq:Qcharge} with
the field-strength tensor \eqref{eq:FijS3} and the corresponding gauge
field \eqref{eq:AiS3} and a simple calculation shows that it reduces
to 
\beq
Q = \frac{1}{4\pi^2}\int_X 
  (\bpsi^\dag\p_i\bpsi)(\p_j\bpsi^\dag\p_k\bpsi)\;
  \d{x}^i\wedge\d{x}^j\wedge\d{x}^k, \label{eq:Q}
\eeq
which is exactly the same as the baryon charge \eqref{eq:B}.
Since the baryon number $B$ \eqref{eq:B} and the Hopf charge
$Q$ \eqref{eq:Q} are given by the same integral expressions, then
$B=Q$ follows.
The final step is to use the fact that preimages of two distinct
regular points on $S^2$ are linked $Q=B$ times under the Hopf
map \eqref{eq:Hopfmap} and hence theorem \ref{thm:1}
follows. \hfill $\square$

\bigskip
Now, if we pick any two regular (constant) points on $S^2$ as
\beq
\bphi_1\in S^2, \qquad
\bphi_2\in S^2,
\eeq
their preimages under the Hopf map composed with $U$, i.e.~$\bphi=H\circ U$, have
linking number $Q=B$.
Since this holds for any two regular points, it also holds for the
following case: Take the two points on $S^2$ to be
\beq
\bphi_1 = H(\bpsi_1,\ol{\bpsi}_1)
= (0,0,-1)^{\rm T}, \qquad
\bphi_2 = H(\bpsi_2,\ol{\bpsi}_2)
= (0,0,1)^{\rm T}, \label{eq:s12}
\eeq
with
\beq
\bpsi_1 = 
\begin{pmatrix}
0\\
1
\end{pmatrix}, \qquad
\bpsi_2 = 
\begin{pmatrix}
1\\
0
\end{pmatrix}.
\label{eq:bpsi_points}
\eeq
\emph{Any} two regular points will have linking number $Q=B$; however,
in order to interpret the preimages of the two points on $S^2$ as two
vortex lines, we further need to require orthogonality
\beq
\bpsi_1^\dag\bpsi_2 = 0,
\eeq
which obviously holds for the two points in
eq.~\eqref{eq:bpsi_points}. 

Clearly it is possible that either both the points \eqref{eq:s12} or
one of them are not regular points.
Since the canonical mapping may not correspond to regular points under
the Hopf map, we propose to rotate the 2-sphere until two regular
points are found: $\bphi^M=M\bphi$ : $X\to S^2$ as
\beq
\bphi^M = M H(\bpsi,\bar{\bpsi}).
\eeq
The most general rotation of the 2-sphere can be done with three Euler
angles and the following parametrization
\begin{align}
M_{\alpha\beta\gamma} &= M_z(\gamma)M_x(\beta)M_z(\alpha), \label{eq:Mabg}\\
M_z(\alpha) &=
\begin{pmatrix}
\cos\alpha & \sin\alpha & 0\\
-\sin\alpha & \cos\alpha & 0\\
0 & 0 & 1
\end{pmatrix},\qquad
M_x(\beta) =
\begin{pmatrix}
1 & 0 & 0\\
0 & \cos\beta & \sin\beta\\
0 & -\sin\beta & \cos\beta
\end{pmatrix}.
\end{align}
A particularly useful rotation brings the north and south poles to the
equator of the 2-sphere:
\beq
M_{0\frac{\pi}{2}\gamma} =
\begin{pmatrix}
\cos\gamma & 0 & \sin\gamma\\
-\sin\gamma & 0 & \cos\gamma\\
0 & -1 & 0
\end{pmatrix},
\eeq
which yields a 1-parameter family of rotations of the north and south
poles to the equator with angle $\gamma\in[0,2\pi)$:
\beq
\bphi_{1,2}^{M_{0\frac{\pi}{2}\gamma}} = 
M_{0\frac{\pi}{2}\gamma}\bphi_{1,2} = \mp
\begin{pmatrix}
\sin\gamma\\
\cos\gamma\\
0
\end{pmatrix},
\label{eq:rotated_bphi_gamma_family}
\eeq
where the upper sign corresponds to $\bphi_1$ and the lower sign
$\bphi_2$.

Another useful rotation is
\beq
M_{0\beta0} = M_x(\beta),
\eeq
which yields a slightly different 1-parameter family of rotations
\beq
\bphi_{1,2}^{M_{0\beta0}} = M_{0\beta0}\bphi_{1,2} = \mp
\begin{pmatrix}
0\\
\sin\beta\\
\cos\beta
\end{pmatrix},
\label{eq:rotated_bphi_beta_family}
\eeq
where again the upper sign corresponds to $\bphi_1$ and the lower sign
$\bphi_2$.

If we now take the parametrization of $\bpsi$
\beq
\bpsi =
\begin{pmatrix}
e^{\i\chi}\cos f\\
e^{\i\vartheta}\sin f
\end{pmatrix}, \label{eq:vortex_parm}
\eeq
we may interpret the two points, $\bpsi_1$ and $\bpsi_2$, of
eq.~\eqref{eq:bpsi_points} as the vortex zeros of the fields $\psi_1$
and $\psi_2$, respectively, see eq.~\eqref{eq:bpsi_def}. 
The composite map $\bphi:X\to S^2$ of eq.~\eqref{eq:vortex_parm} thus
reads 
\beq
\bphi =
\begin{pmatrix}
\sin 2f \cos(\vartheta - \chi)\\
\sin 2f \sin(\vartheta - \chi)\\
\cos 2f
\end{pmatrix},
\eeq
from which it is clear that the two points $\bphi_{1,2}\in S^2$ of
eq.~\eqref{eq:s12} indeed are independent of $\vartheta$ and $\chi$ as
they correspond to $f=\frac\pi2$ and $f=0$, respectively. 
These two vortex zeroes are canonically mapped to the south and north
poles, respectively.
Using now the rotated map $\bphi^{M_{0\frac\pi2\gamma}}$ of
eq.~\eqref{eq:rotated_bphi_gamma_family}, the
vortex \eqref{eq:vortex_parm} is mapped to
\beq
\bphi^{M_{0\frac\pi2\gamma}} =
\begin{pmatrix}
\sin\gamma\cos2f + \cos\gamma\sin2f\cos(\vartheta-\chi)\\
\cos\gamma\cos2f - \sin\gamma\sin2f\cos(\vartheta-\chi)\\
-\sin2f\sin(\vartheta-\chi)
\end{pmatrix},
\eeq
which at $f=\frac\pi2,0$ equals
eq.~\eqref{eq:rotated_bphi_gamma_family} by construction.

\begin{corollary}\label{crl:1}
Two vortex lines (zeros), $\bpsi_1$ and $\bpsi_2$ of
$\bpsi\in X=\mathbb{R}^3\cup\{\infty\}$
are mapped to two distinct points on $S^2$ under the Hopf map
$H\circ U:X\to S^2$ and hence their preimages in $X$ are linked $Q=B$
times due to theorem \ref{thm:1}. 
\end{corollary}

We may take a map $U:X\to N$ with topological degree $B$, project it
onto $S^2$ with $H$ and select two regular points under the latter
mapping
\beq
\bphi_{1,2}^{M_{0\frac\pi2\gamma}} = H\circ U,
\eeq
where we have performed a rotation using
eq.~\eqref{eq:rotated_bphi_gamma_family} and chosen an appropriate
value for $\gamma$ such that the mapping is regular.
Now due to the Corollary \ref{crl:1}, we can follow the way back to
$X$ with the inverse mappings and interpret the two points as vortex
lines
\beq
\bx_{1,2}(\tau,\ell) = (H\circ U)^{-1}\left(\bphi_{1,2}^{M_{0\frac\pi2\gamma}}\right),
\eeq
which yields two vortex lines with some parametrization $\tau$ and we
have included an index $\ell$ in case the preimages separate into several
clusters. 

We are now ready to make the following conjecture.

\begin{conjecture}\label{cjt:1}
A map $U:\mathbb{R}^3\cup\{\infty\}=X\to S^3$ having degree
$B$ \eqref{eq:B} can be interpreted as two vortices in $\psi_1$ and
$\psi_2$ of $\bpsi\in S^3$ which in each cluster topologically have
winding numbers $p_\ell$ and $q_\ell$, respectively. Then due to
Corollary \ref{crl:1}, the linking number $Q$ \eqref{eq:Q} is
$\sum_{\ell}p_{\ell}q_{\ell}$ and due to theorem \ref{thm:1},
$B=Q=\sum_{\ell}p_{\ell}q_{\ell}$. 
\end{conjecture}

\subsection{The rational map}

We will now consider $U$ to be in a class of maps, where it is a
radial suspension in $\mathbb{R}^3$ and the tangent directions are
described by rational maps between Riemann spheres. 
The rational map Ansatz is given by
\beq
U = \exp\left(\i f(r) \mathbf{n}\cdot\mathbf{\tau}\right),
\label{eq:UR}
\eeq
with
\beq
\mathbf{n} =
\left(
\frac{R+\ol{R}}{1+|R|^2},
\frac{\i(\ol{R}-R)}{1+|R|^2},
\frac{1-|R|^2}{1+|R|^2}
\right),
\eeq
where $R=R(z)$ is a holomorphic function of the Riemann sphere
coordinate $z=e^{\i\phi}\tan\tfrac{\theta}{2}$ and $(r,\theta,\phi)$ are
standard spherical coordinates in $\mathbb{R}^3$.

Using eq.~\eqref{eq:Udef}, we get
\beq
\bpsi = \frac{1}{1+|R|^2}
\begin{pmatrix}
e^{\i f} + |R|^2 e^{-\i f}\\
\i 2R\sin f
\end{pmatrix},
\eeq
which we map to the 2-sphere using $H$ of eq.~\eqref{eq:Hopfmap},
yielding 
\beq
\bphi = \frac{1}{(1+|R|^2)^2}
\begin{pmatrix}
-2\Im(R)(1+|R|^2)\sin 2f + 4\Re(R)(1-|R|^2)\sin^2f\\
2\Re(R)(1+|R|^2)\sin 2f + 4\Im(R)(1-|R|^2)\sin^2f\\
4|R|^2\cos 2f + (1-|R|^2)^2
\end{pmatrix}. \label{eq:HopfR}
\eeq
It is easy to check that the above $\bphi$ is a real-valued 3-vector
of unit norm, thus living on $S^2$.
One can also readily verify that $f=\frac\pi2,0$ correspond to
$\bphi_{1,2}$, respectively.

\section{Examples}\label{sec:examples}

\subsection{Toroidal vortex}

Let us consider a simple example inspired by
ref.~\cite{Gudnason:2016yix,Gudnason:2014gla,Gudnason:2014hsa,Gudnason:2014jga,Gudnason:2018oyx}
where a vortex ring is twisted $P$ times, yielding baryon number $P$: 
\beq
\bpsi =
\begin{pmatrix}
\cos f(r) + \i\sin f(r)\cos\theta\\
\sin f(r)\sin\theta e^{\i P\phi}
\end{pmatrix}.
\eeq
\begin{figure}[!htp]
\begin{center}
\includegraphics[width=0.5\linewidth]{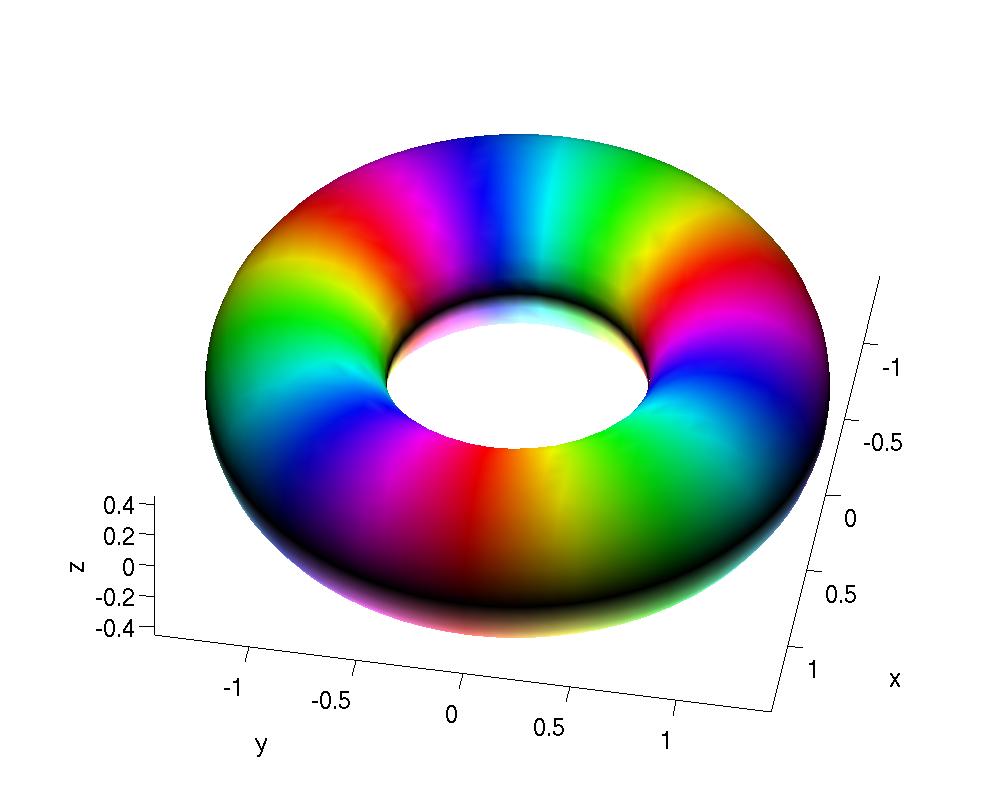}
\caption{Vortex ring with $P=3$ twists. Figure taken from
ref.~\cite{Gudnason:2014jga}.} 
\label{fig:tvtx}
\end{center}
\end{figure}
The energy functional that gives rise to toroidal vortices is given by 
\cite{Gudnason:2016yix}
\beq
E[\bpsi] = \|\d\bpsi\|_{L^2(X)}^2
+ \frac14\|\bpsi^*\d{\mu}\|_{L^2(X)}^2
+ \int_X*V(\bpsi), \qquad
V(\bpsi) = \frac{m^2}{2}(1 - |\psi_1|^2), \label{eq:toroidal}
\eeq
where $\mu$ is the Maurer-Cartan form on $N$ (SU(2)) and the second term
is the norm-squared of the pullback of the exterior derivative of the
Maurer-Cartan form on $N$ by $\bpsi$.
$m$ is a positive constant which must be large enough $m>m_{\rm
crit}$.
Finally $*$ denotes the Hodge dual such that $*1$ gives the volume
form (and in this case on $X$). 
The baryon charge density isosurface is shown in fig.~\ref{fig:tvtx},
which is taken from ref.~\cite{Gudnason:2014jga} where further details
can be found.
It is easy to check that the topological degree \eqref{eq:B} is given
by 
\begin{align}
B &= -\frac{P}{\pi}\int_0^\pi \sin\theta\,\d{\theta}
  \int_0^\infty \sin^2f(r) \p_rf(r)\,\d{r} \non
  &= -\frac{P}{2\pi}\left[2f(r) - \sin 2f(r)\right]_{f(0)}^{f(\infty)} \non
  &= P,
\end{align}
where we have used the boundary conditions $f(0)=\pi$ and
$f(\infty)=0$.

Under the map \eqref{eq:Hopfmap} we have
\beq
\bphi =
\begin{pmatrix}
\sin\theta\cos P\phi\sin 2f(r) + \sin 2\theta\sin P\phi\sin^2f(r)\\
\sin\theta\sin P\phi\sin 2f(r) - \sin 2\theta\cos P\phi\sin^2f(r)\\
\sin^2\theta\cos 2f(r) + \cos^2\theta
\end{pmatrix}.
\eeq
An obvious choice would be to pick the two points $\bphi_{1,2}$ of
eq.~\eqref{eq:s12} on the 2-sphere, yielding
\begin{align}
\cos^2f = \cos^2\theta &= 0, \qquad (\bphi=\bphi_1, 
\; \psi_1=0)\\
\sin^2 f\sin^2\theta &= 0, \qquad (\bphi=\bphi_2,\; \psi_2=0).
\end{align}
Let us start with the latter equation; $\sin f=0$ corresponds to the vacuum
at $r\to\infty$ and the origin where $f=\pi$. Hence in the interior of
$\mathbb{R}^3\backslash\{0\}$, $\sin f\neq 0$ and thus $\sin\theta=0$
corresponds to the $x^3$ axis.
We call it a ``vacuum vortex,'' specified by $\psi_2=0$.
On $X\simeq S^3$ this is topologically a circle ($S^1$) going from the
north pole of the 3-sphere (the vacuum) from $x^3=-\infty$ to the
origin $\bx=\mathbf{0}$ which is the south pole of the 3-sphere, and
then towards $x^3\to+\infty$ back to north pole.
The former equation has two conditions yielding $\theta=\frac\pi2$ and
$f(r)=\frac\pi2$, which is a circle in the $(x^1,x^2)$-plane,
representing a ring-shaped ``physical'' vortex specified by
$\psi_1=0$. 
This obviously yields a linking number equal to 1.
Although this is a natural interpretation of where the two vortices
might be in this field configuration, $\bphi_2$ is not a regular point
of the mapping when $P>1$.

\begin{figure}[!htp]
\begin{center}
\mbox{\subfloat[$\beta=0$]{\includegraphics[height=0.4\linewidth]{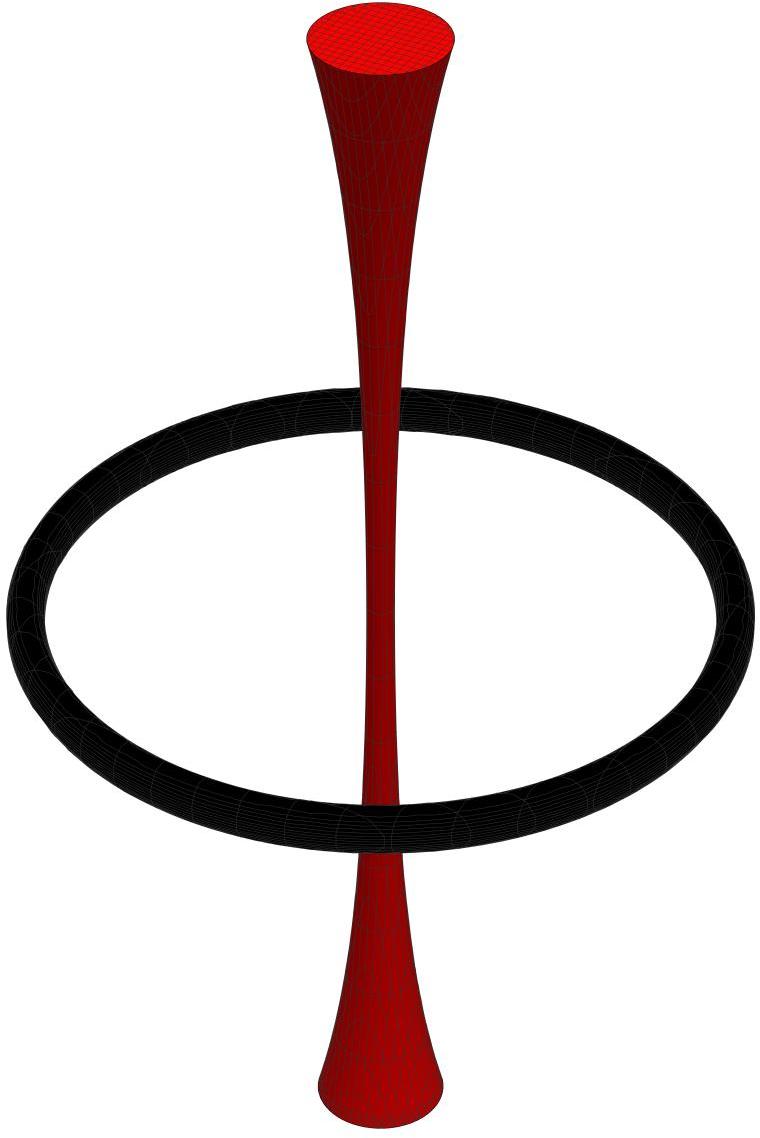}}\ \
\subfloat[$\beta=\frac\pi6$]{\includegraphics[height=0.3\linewidth]{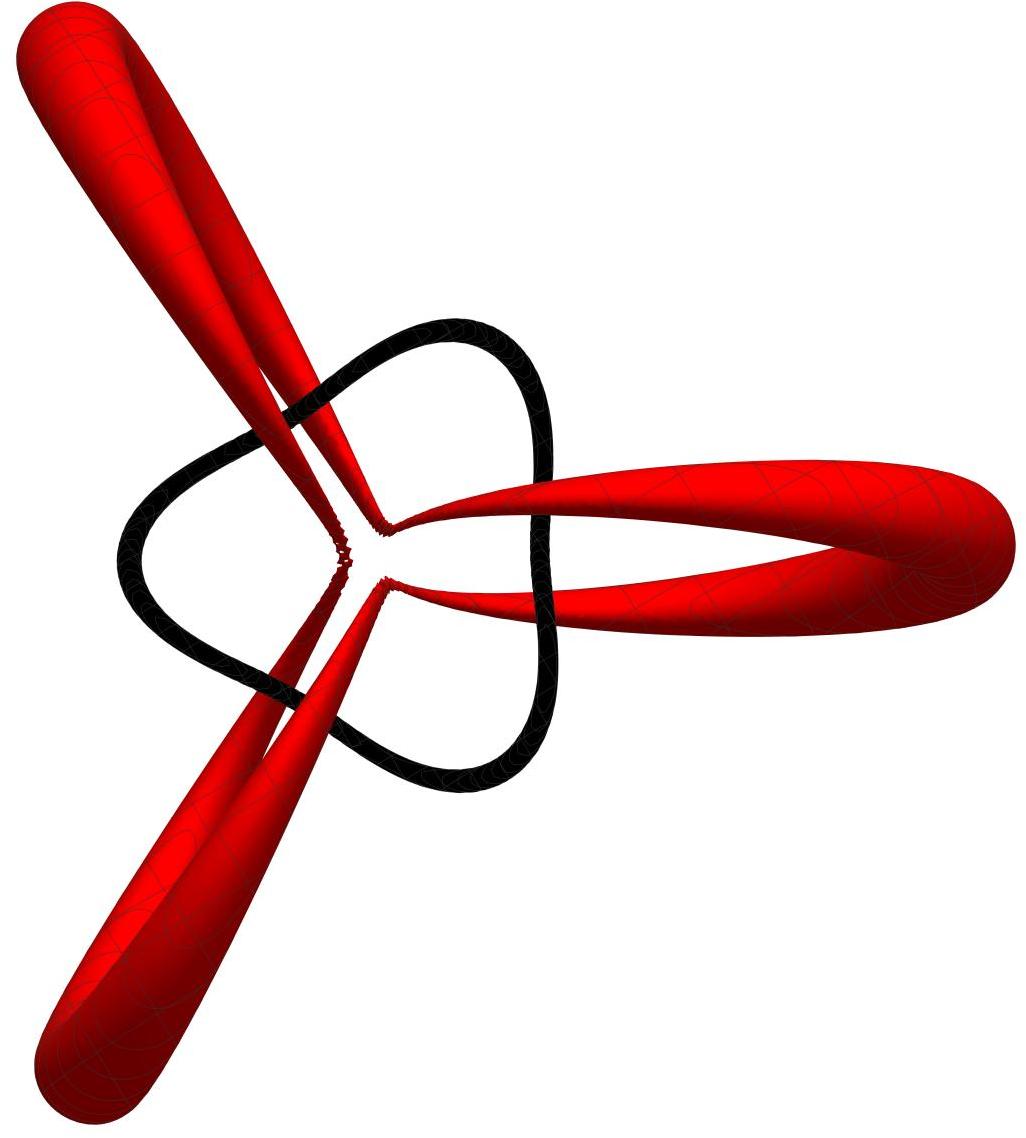}}\ \
\subfloat[$\beta=\frac\pi2$]{\includegraphics[height=0.28\linewidth]{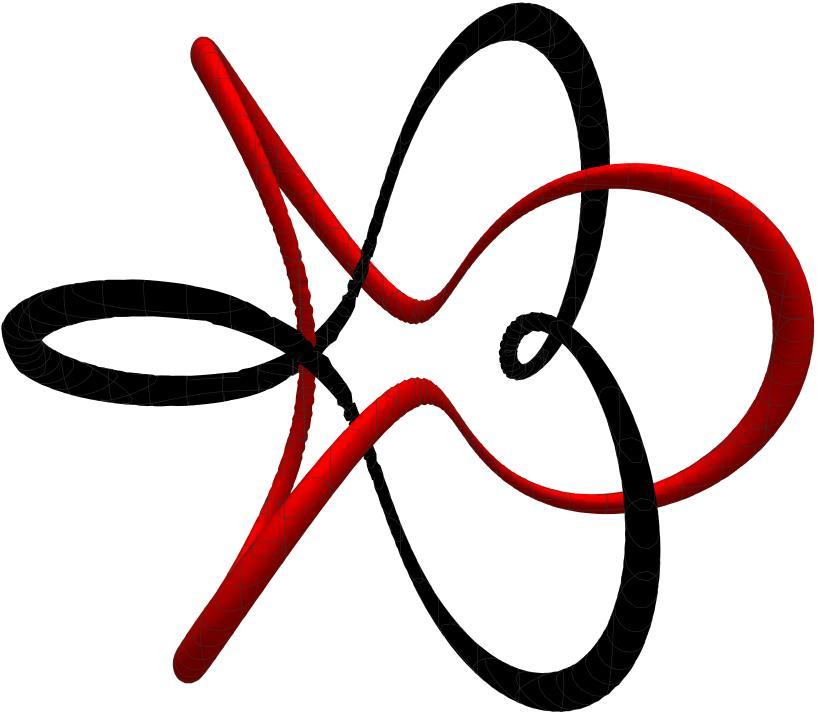}}}
\caption{The toroidal vortex (black) with $P=3$ twists and its vacuum
vortex (red) at rotation angles $\beta=0,\frac\pi6,\frac\pi2$.
The angle $\beta=0$ corresponds to no rotation and shows the
degeneracy of the vacuum vortex which is due to the north pole on the
2-sphere not being a regular point under the Hopf map. }
\label{fig:tvtx_beta}
\end{center}
\end{figure}

In order to move away from the point where the $P$ vortices linking
the ``vacuum vortex'' are degenerate, we pick two points on the
2-sphere after a rotation by an angle $\beta$:
\beq
\bphi_{1,2}^{M_{0\beta0}},
\eeq
with $M_{\alpha\beta\gamma}$ given by eq.~\eqref{eq:Mabg} and
$\bphi_{1,2}$ given by eq.~\eqref{eq:s12}.
The expression is not particularly illuminating, so we will just plot
the preimages of the two points on the rotated 2-sphere in
fig.~\ref{fig:tvtx_beta}.

Plotting the preimage of one of the points on the 2-sphere amounts to
finding the solutions to the inverse map
\beq
\mathbf{x} = \bphi^{-1}(\bphi_a),
\eeq
with $\bphi_a$ a chosen point on $S^2\ni\bphi_a=\bphi(\mathbf{x})$.
In practice, the solution to this problem is a hairline and not easy
to see on a 3-dimensional graph, so we plot instead a surface that
corresponds to 1\% of the neighborhood around $\bphi_a$. 
In particular, if we want to plot $\mathbf{x}=\bphi^{-1}((0,0,-1))$,
we instead plot the surface $\bphi^{-1}\big((a,b,-\sqrt{1-a^2-b^2})\big)$ with
$\sqrt{a^2+b^2}=0.01$.

Fig.~\ref{fig:tvtx_beta}(a) shows the unrotated degenerate case, where
the vacuum vortex with winding number 3 is coincident -- this thus
corresponds to a point on the 2-sphere which is not regular under the
Hopf map \eqref{eq:Hopfmap}.
In fig.~\ref{fig:tvtx_beta}(b) we have increased $\beta$ to
$\beta=\frac\pi6$ and we have moved away from the degeneracy of the
vacuum vortex. Now we can clearly see that the vortex ring, which is
the black circle depicted in fig.~\eqref{fig:tvtx_beta}(a), is linked
3 times with the vacuum vortex (red).
Note that both preimages are themselves not knots, but indeed
unknots.

A comment in store is about the black line, i.e.~the vortex ring
itself in fig.~\ref{fig:tvtx_beta}. In fig.~\ref{fig:tvtx_beta}(a) the
preimage shows the center of the vortex and what one normally would
associate with the position of the vortex; unfortunately the antipodal
point on the 2-sphere under the Hopf map is not regular, as mentioned
above.
Once we rotate the two points, $\bphi_{1,2}$, keeping them antipodal
on the 2-sphere, we also move the vortex point itself and the preimage 
runs $P$ times around the vortex center line on fixed level sets of
the vortex field.
At the $\beta=\frac\pi2$ rotation, we have rotated all the way to the
equator, which corresponds to $\gamma=0$ of
eq.~\eqref{eq:rotated_bphi_gamma_family}.
Here the vortex line and the vacuum vortex becomes identical, except
that one is rotated by $\pi/P=\pi/3$ with respect to the other. 

This example thus confirms conjecture \ref{cjt:1} with the vortex ring
having $q=1$ and the vacuum vortex having $p=3$, yielding $B=Q=pq=3$.

Before moving on to the next example, let us make one more comment.
The energy \eqref{eq:toroidal} is an example where the potential term
is asymmetric in $\psi_1$ and $\psi_2$, so the physical vortex zeros
correspond to $\psi_1=0$ and the vacuum vortex zeros to $\psi_2=0$.
Instead, we could consider the potential term which is symmetric 
in $\psi_1$ and $\psi_2$
\cite{Gudnason:2014gla,Gudnason:2014hsa,Gudnason:2014jga,Gudnason:2018oyx}
\beq
V(\bpsi) = \pm\frac{m^2}{8}\big[1 - (\bpsi^\dag \sigma_3 \bpsi)^2\big] 
= \pm \frac{m^2}{2} |\psi_1|^2 |\psi_2|^2. \label{eq:toroidal2}
\eeq
For the positive sign, there are two vacua: $\bpsi=(e^{\i\alpha},0)$
and $(0,e^{\i\alpha})$ 
\cite{Gudnason:2014gla,Gudnason:2014hsa,Gudnason:2014jga,Gudnason:2018oyx},
while for the negative sign the vacuum is: $S^1\times S^1$, 
$|\psi_1|^2=|\psi_2|^2=1/2$. 
In the former case, the situation is similar to that of the
potential \eqref{eq:toroidal} admitting physical vortex zeros and
vacuum vortex zeros, while in the latter case both zeros can be
physical vortices. 
This potential is motivated by two-component Bose-Einstein condensates
(BEC) \cite{Kasamatsu}, and we called the model the BEC-Skyrme model,
see Appendix A of ref.~\cite{Gudnason:2014hsa} for a more precise
correspondence.  
In fact, a Skyrmion in two-component BECs was constructed as 
a link of two kinds of vortices 
\cite{Ruostekoski:2001fc,Battye:2001ec,Nitta:2012hy}.

\subsection{Rational map Skyrmions}

We will now illustrate theorem \ref{thm:1} and conjecture \ref{cjt:1}
using the rational map approximation to Skyrmion
solutions \cite{Houghton:1997kg}.
The energy functional is now simply given by
\beq
E[\bpsi] = \|\d\bpsi\|_{L^2(X)}^2
+ \frac14\|\bpsi^*\d{\mu}\|_{L^2(X)}^2,
\label{eq:massless_Skyrme}
\eeq
see the previous subsection for an explanation.
Inserting the rational map Ansatz \eqref{eq:UR} yields
\beq
E[f,R] = \int_0^\infty\left(f^{'2} + 2B\sin^2f(f^{'2}+1)
+ \mathcal{I}[R]\frac{\sin^4f}{r^4}\right) r^2\; \d{r},
\eeq
with
\beq
\mathcal{I}[R] = \frac{1}{4\pi}\|R^*\Omega_N\|_{L^2(S^2)}^2,
\label{eq:calI}
\eeq
where the only way the rational map enters the energy functional is
through this integral which is the norm-squared of the pullback of the
area form on $N$ by $R$, and $B$ is the degree of the rational map $R$.

We will thus utilize the map \eqref{eq:HopfR} with $R(z)$ being the
rational map of degree $B$.
Plotting the points \eqref{eq:s12} corresponds to a vortex (which is
the antivacuum of the Skyrmion) and the vacuum vortex (which contains the
vacuum).
In all cases, except the $B=1$ case, the vacuum vortex does not
correspond to a regular point under the map \eqref{eq:HopfR} and hence
the preimages degenerate, making it impossible to count the linking
number -- which indeed is in accord with theorem \ref{thm:1}.
For certain $B$, even the vortex (the antivacuum) does not correspond
to a regular point under the mapping. 
Therefore, we turn to (two) antipodal points on the 2-sphere, which do 
correspond to regular points under the map \eqref{eq:HopfR} by
rotating the 2-sphere using
eqs.~\eqref{eq:rotated_bphi_gamma_family}, \eqref{eq:rotated_bphi_beta_family}
and the 
linking numbers thus exactly equal the baryon numbers of the
solitons. 

In order to facilitate the visualization of the preimages of the
soliton solutions in terms of Skyrmion maps, it will prove helpful to
plot a fixed level set of the baryon charge density so as to get a
frame of reference for the preimages.
The baryon charge density is given by
\beq
\mathcal{B} = *\bpsi^*\Omega_N,
\eeq
which is a 0-form (scalar quantity) and is calculated as the Hodge
dual on $X$ of the pullback of the normalized volume form on $N$ by
the map $\bpsi$.

We are now ready to present the results of various preimages of the
points $\bphi_{1,2}^{M_{0\frac\pi2\gamma}}$ and
$\bphi_{1,2}^{M_{0\beta0}}$ for rational map Skyrmions with
$B=1,2,\ldots,8$.
We will display the degenerate points $\bphi_{1,2}$ just for reference.

\begin{figure}[!htp]
\begin{center}
\mbox{\subfloat[$\bphi_{1,2}$]{\includegraphics[scale=0.11]{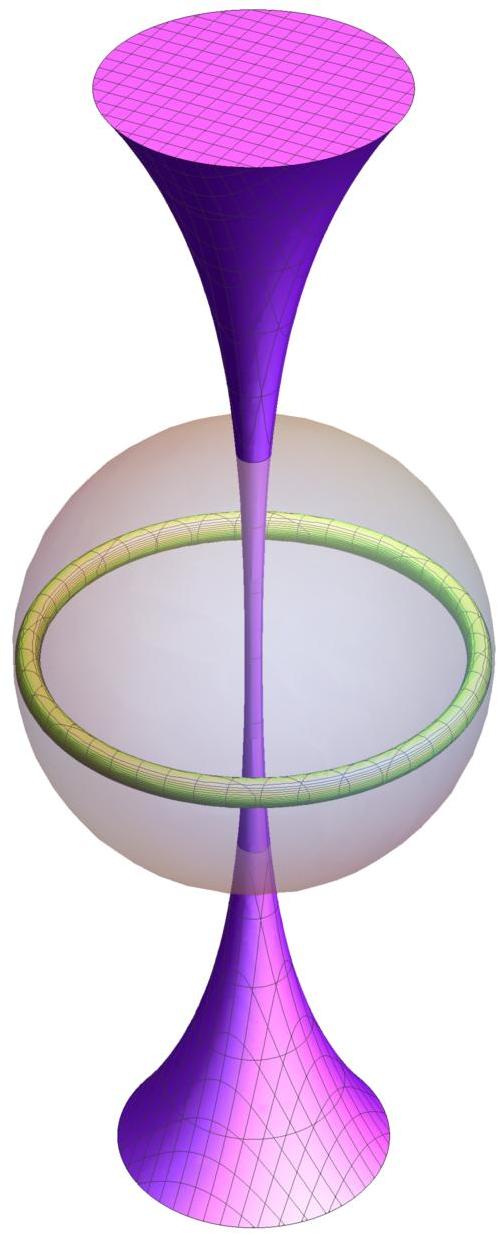}}
\subfloat[$\bphi_{1,2}^{M_{0\frac\pi2\frac\pi4}}$]{\includegraphics[scale=0.12]{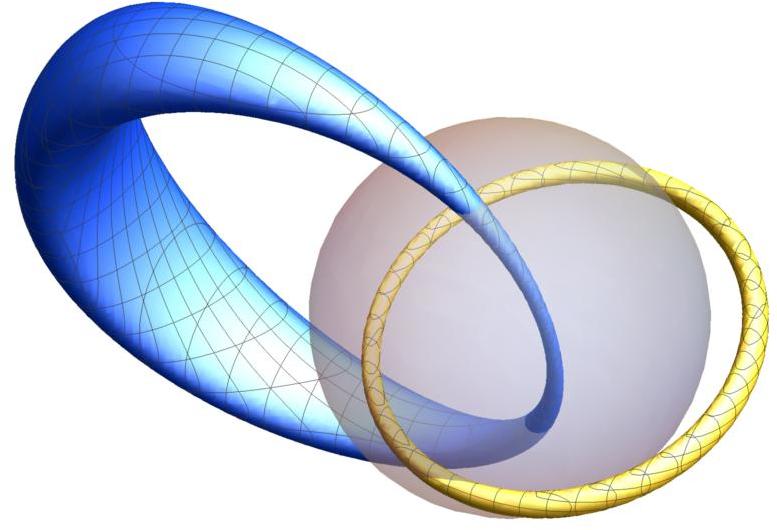}}}
\caption{Links for the $B=1$ Skyrmion.
(a) A link between the vortex ring (yellow) and the vacuum vortex
(magenta). (b) A link between the vortex 
(yellow) and the vacuum vortex (blue) which is a closed loop.
The gray isosurface is the baryon charge density illustrating the
shape of the Skyrmion.
}
\label{fig:RMB1}
\end{center}
\end{figure}

The rational map Skyrmion with topological degree 1 is given by the
spherically symmetric rational map \cite{Houghton:1997kg}
\beq
R_1(z) = z.
\eeq
Fig.~\ref{fig:RMB1} shows the preimages of $\bphi_{1,2}$ and again
after a rotation using the rotation matrix $M_{0\frac\pi2\frac\pi4}$
of eq.~\eqref{eq:rotated_bphi_gamma_family} has been applied.
In this case, and only in this case, $\bphi_{1,2}$ are regular points
under the mapping \eqref{eq:HopfR}.
The vacuum vortex (magenta) in fig.~\ref{fig:RMB1}(a) goes from
$x^3=-\infty$ to $x^3=\infty$, which are identified by the one-point
compactification and hence it is a vortex ring, linking the other
vortex ring (yellow) exactly once, as expected.

In order to see what happens to the preimages once the rotation of the
2-sphere has been applied, we show
$\bphi_{1,2}^{M_{0\frac\pi2\frac\pi4}}$ in fig.~\ref{fig:RMB1}(b).
The two points are still antipodal on the 2-sphere in order to lend
the interpretation as ``vortices,'' but it is clear that the vortex
(yellow) is slightly shifted and the vacuum vortex (blue) is now
closing in the bulk of $\mathbb{R}^3$.
Topologically it is the same thing of course and since both points are
regular, they both give linking number $Q=B=1$ as theorem \ref{thm:1}
states and the interpretation as vortex links according to
conjecture \ref{cjt:1} is also clear.

\begin{figure}[!htp]
\begin{center}
\mbox{\subfloat[$\bphi_{1,2}$]{\includegraphics[scale=0.08]{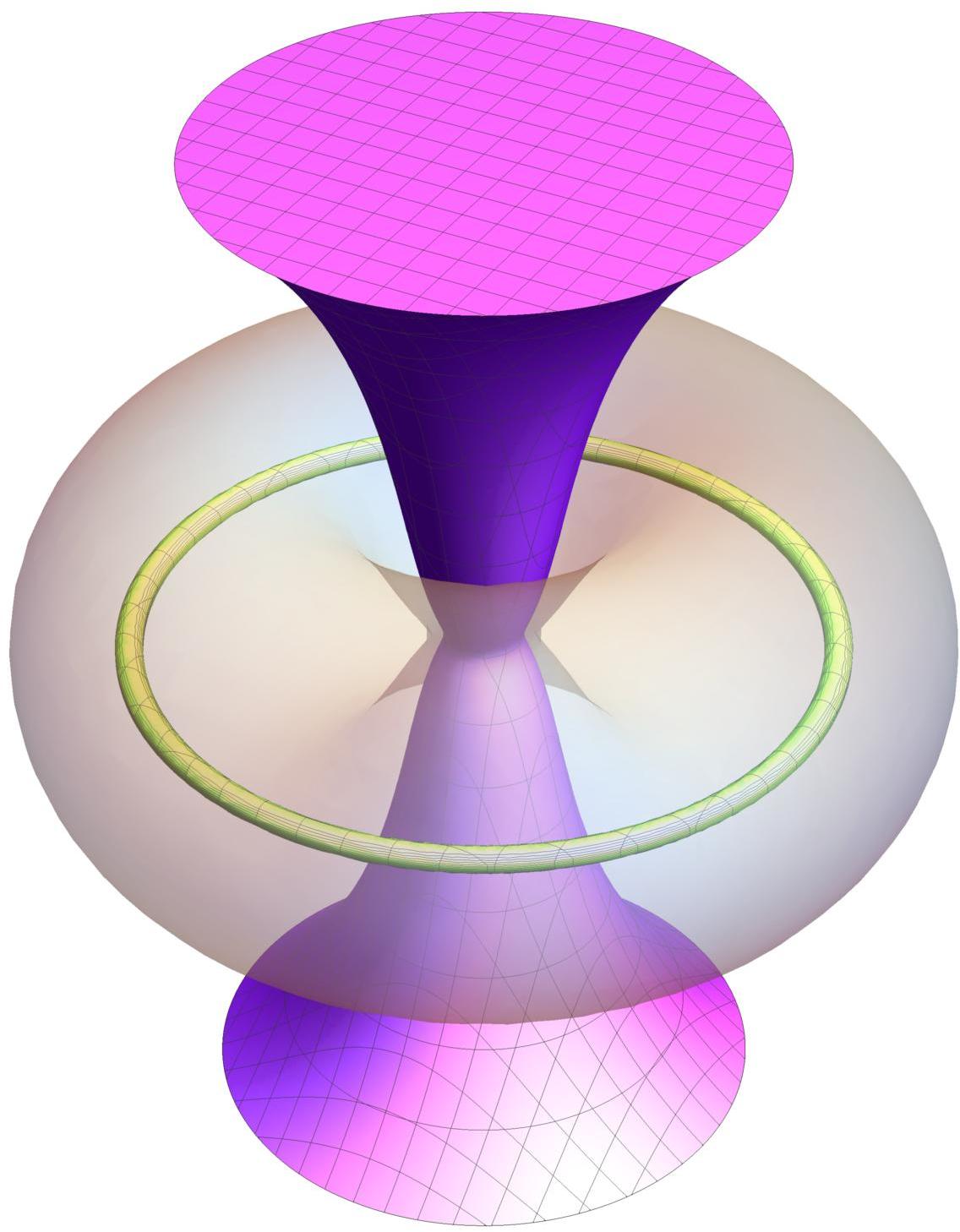}}
\subfloat[$\bphi_{1,2}^{M_{0\frac\pi60}}$]{\includegraphics[scale=0.14]{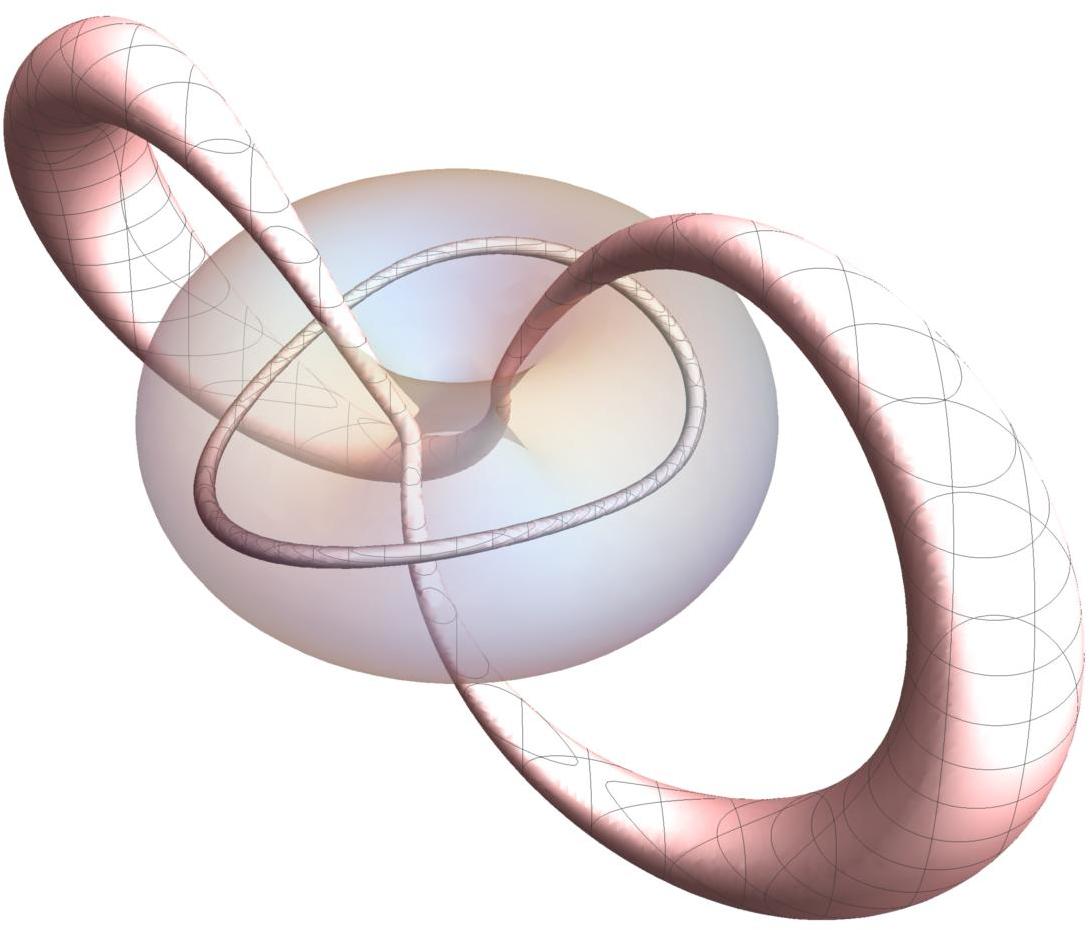}}
\subfloat[$\bphi_{1,2}^{M_{0\frac\pi20}}$]{\includegraphics[scale=0.12]{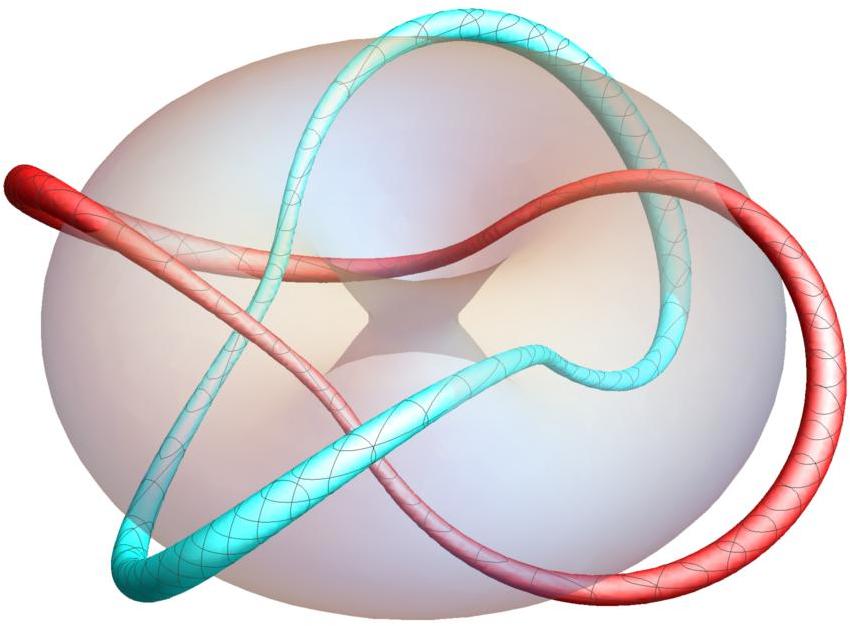}}}
\caption{Links for the $B=2$ Skyrmion.
(a) A link between the vortex ring (yellow) and the vacuum vortex
(magenta) which is degenerate. (b,c) Nondegenerate links between the
vortex and vacuum vortex, which are both closed loops.
The gray isosurface is the baryon charge density illustrating the
shape of the Skyrmion.
}
\label{fig:RMB2}
\end{center}
\end{figure}

The rational map Skyrmion with topological degree 2 is given by the
axially symmetric rational map \cite{Houghton:1997kg}
\beq
R_2(z) = z^2.
\eeq
Fig.~\ref{fig:RMB2} shows preimages of $\bphi_{1,2}$ and again after a
rotation by $\beta=\frac\pi6$ and $\beta=\frac\pi2$.
The vacuum vortex (magenta) in fig.~\ref{fig:RMB2}(a) is degenerate
and this is because the point on the 2-sphere is not a regular point
under the mapping \eqref{eq:HopfR}, as mentioned already.
Rotating the points, keeping them mutually antipodal, the preimages of
fig.~\ref{fig:RMB2}(b) are perfectly linked twice and in
fig.~\ref{fig:RMB2}(c) the vacuum vortex becomes identical with the
vortex, albeit with a $\pi/2$ rotation with respect to the latter.
This example confirms conjecture \ref{cjt:1} with the vortex ring
having $q=1$ and the vacuum vortex having $p=2$, yielding $B=Q=pq=2$.

\begin{figure}[!htp]
\begin{center}
\mbox{\subfloat[$\bphi_{1,2}$]{\includegraphics[scale=0.085]{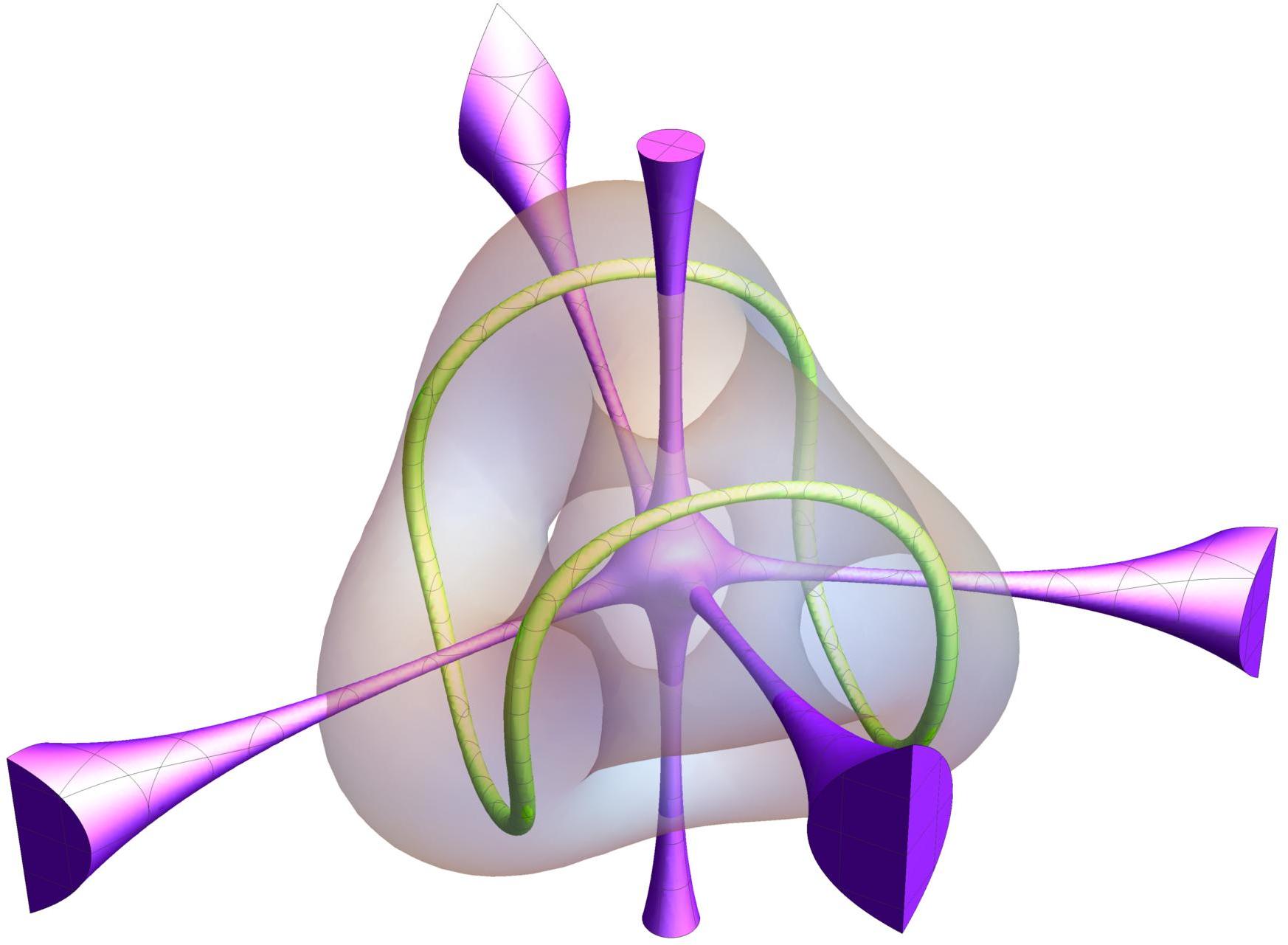}}
\subfloat[$\bphi_{1,2}^{M_{0\frac\pi60}}$]{\includegraphics[scale=0.13]{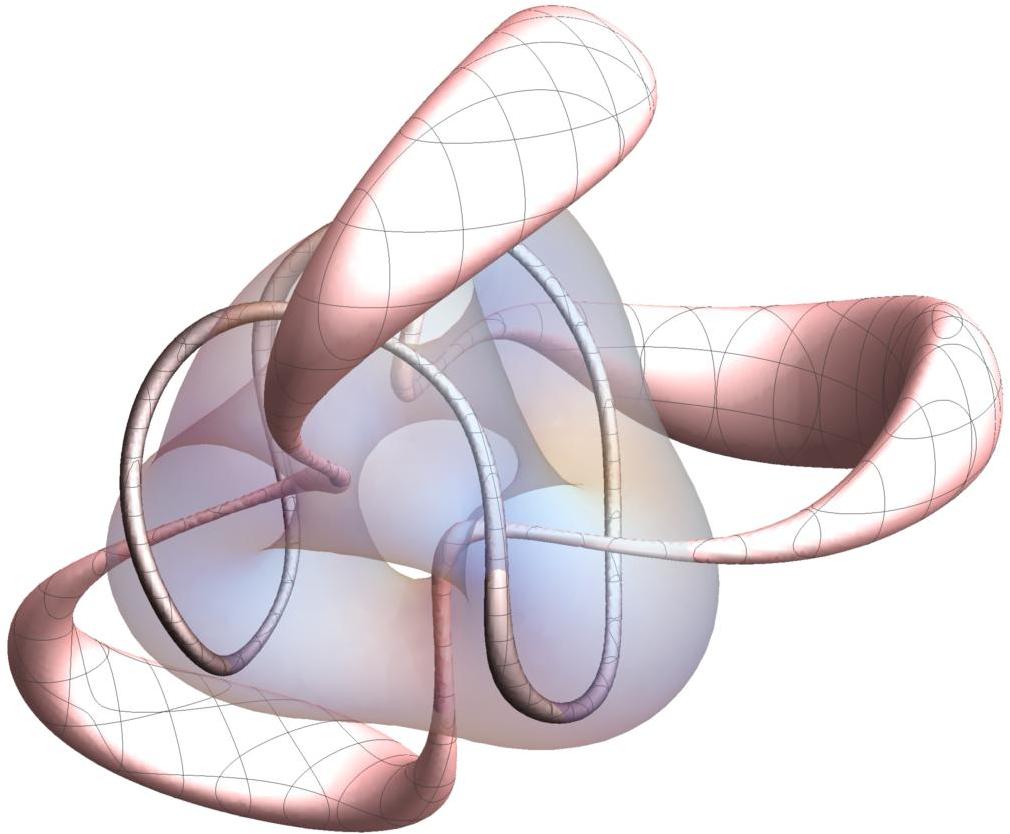}}}
\mbox{\subfloat[$\bphi_{1,2}^{M_{0\frac\pi2\frac\pi4}}$]{\includegraphics[scale=0.13]{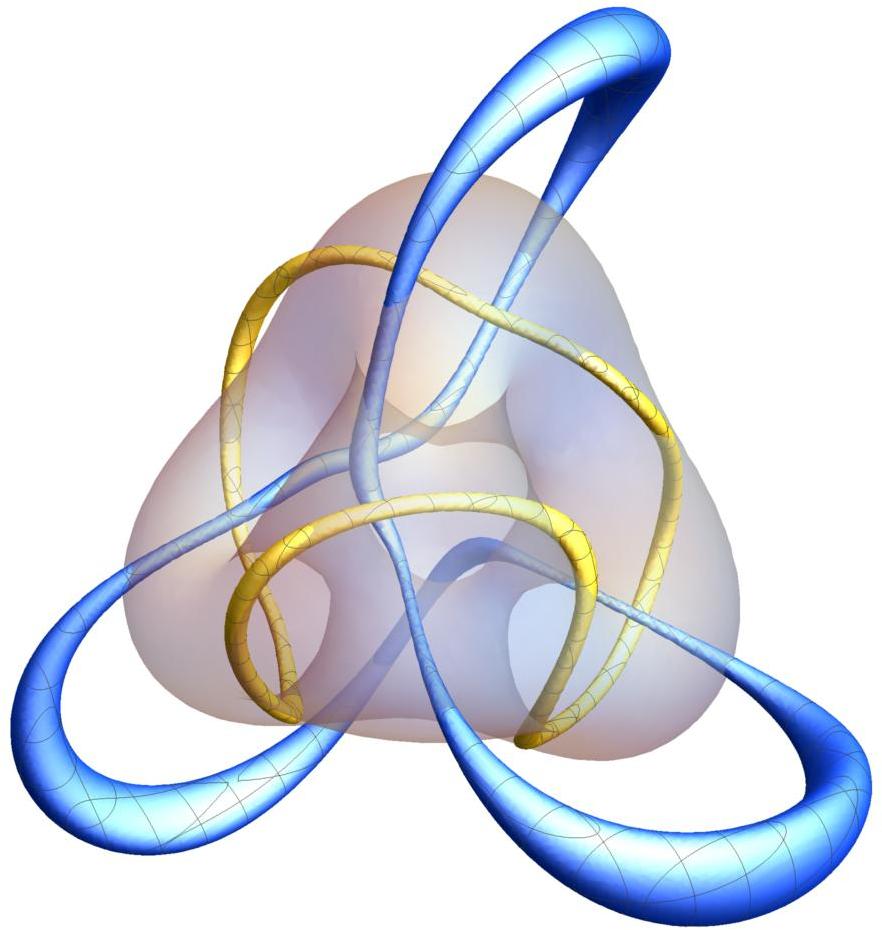}}}
\caption{Links for the $B=3$ Skyrmion.
(a) A link between the vortex ring (yellow) and the vacuum vortex
(magenta) which is degenerate. (b,c) Nondegenerate links between the
vortex and vacuum vortex, which are both closed loops.
The gray isosurface is the baryon charge density illustrating the
shape of the Skyrmion.
}
\label{fig:RMB3}
\end{center}
\end{figure}

The next soliton is the rational map Skyrmion of topological degree
3. The rational map is given by \cite{Houghton:1997kg}
\beq
R_3(z) = \frac{\i\sqrt{3}z^2 - 1}{z(z^2 - \i\sqrt{3})},
\eeq
and possesses tetrahedral symmetry.
Fig.~\ref{fig:RMB3} shows preimages of $\bphi_{1,2}$ as well as two
rotations by $\beta=\frac\pi6$ and by $\gamma=\frac\pi4$.
The vacuum vortex (magenta) in fig.~\ref{fig:RMB3}(a) is still
degenerate as mentioned above.
There is now evidence for the vacuum vortex of rational map Skyrmions
to be $B$ intersecting (infinite) lines coming from and returning to
$\p\mathbb{R}^3$.
After a suitable rotation as shown in fig.~\ref{fig:RMB3}(b,c) the
linking number of two antipodal points on the 2-sphere is now equal to
three, as promised.
This example confirms conjecture \ref{cjt:1} with the vortex ring
having $q=1$ and the vacuum vortex having $p=3$, yielding $B=Q=pq=3$.

\begin{figure}[!htp]
\begin{center}
\mbox{\subfloat[$\bphi_{1,2}$]{\includegraphics[scale=0.095]{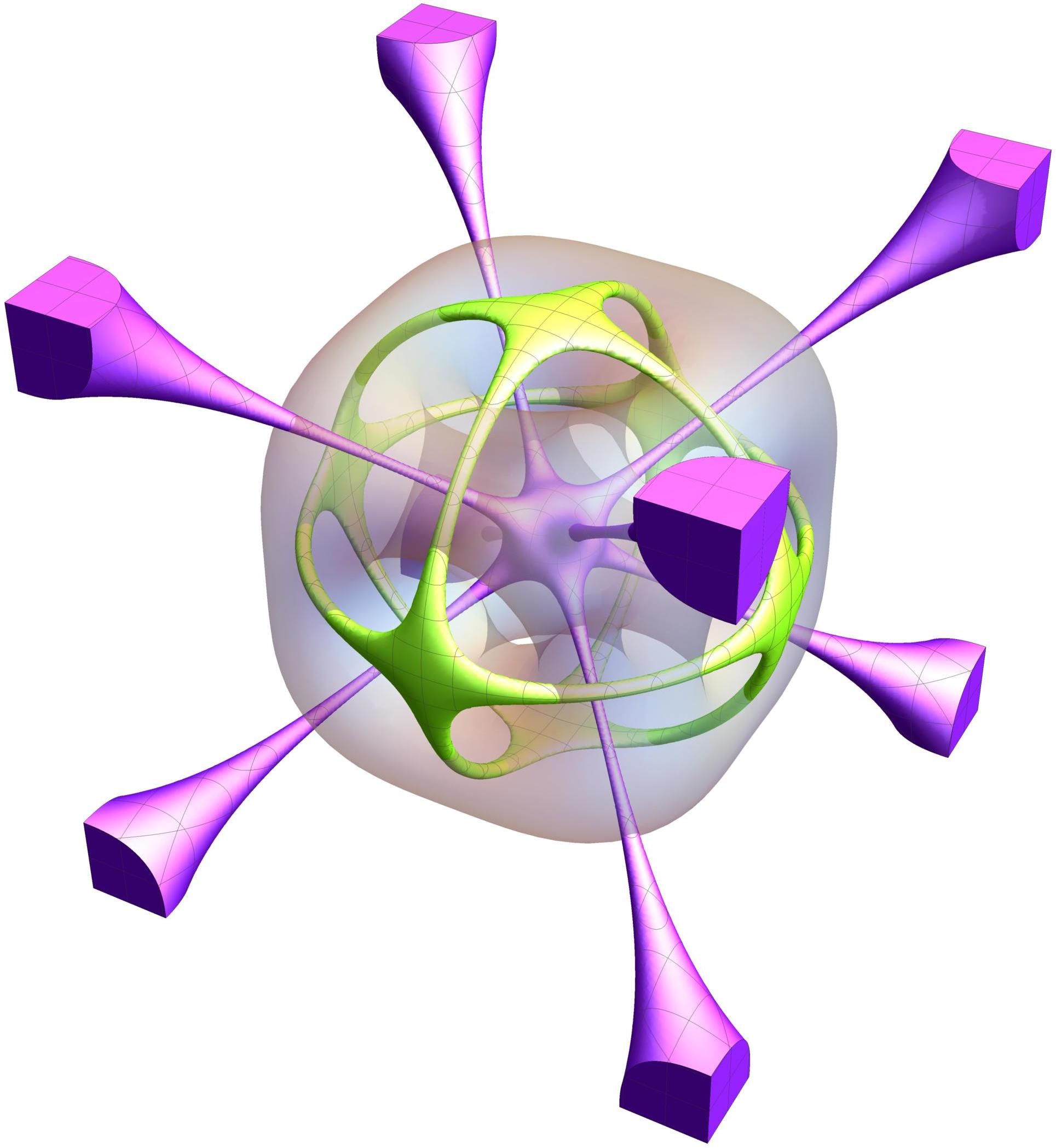}}
\subfloat[$\bphi_{1,2}^{M_{0\frac\pi60}}$]{\includegraphics[scale=0.14]{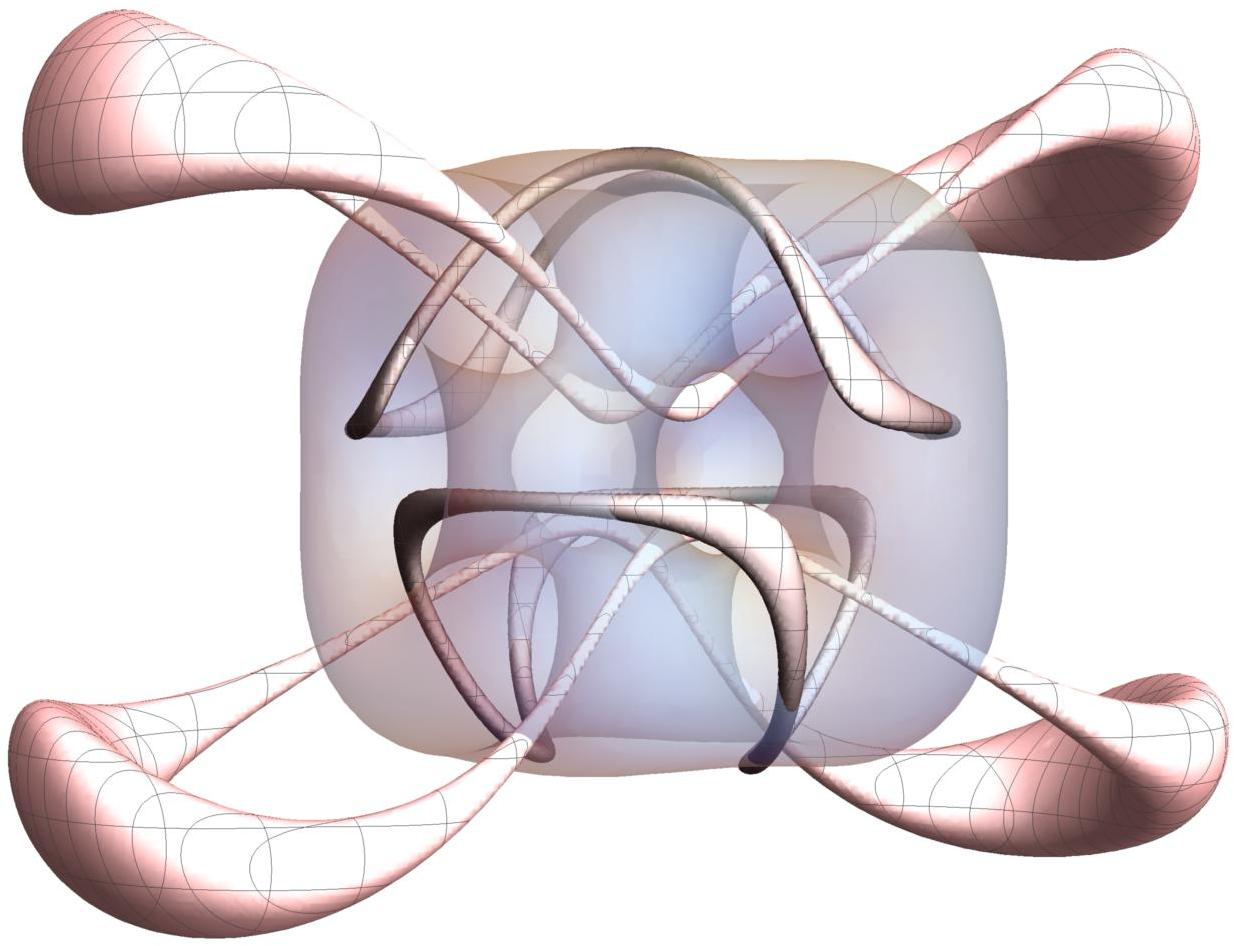}}}
\mbox{\subfloat[$\bphi_{1,2}^{M_{0\frac{3\pi}{2}0}}$]{\includegraphics[scale=0.108]{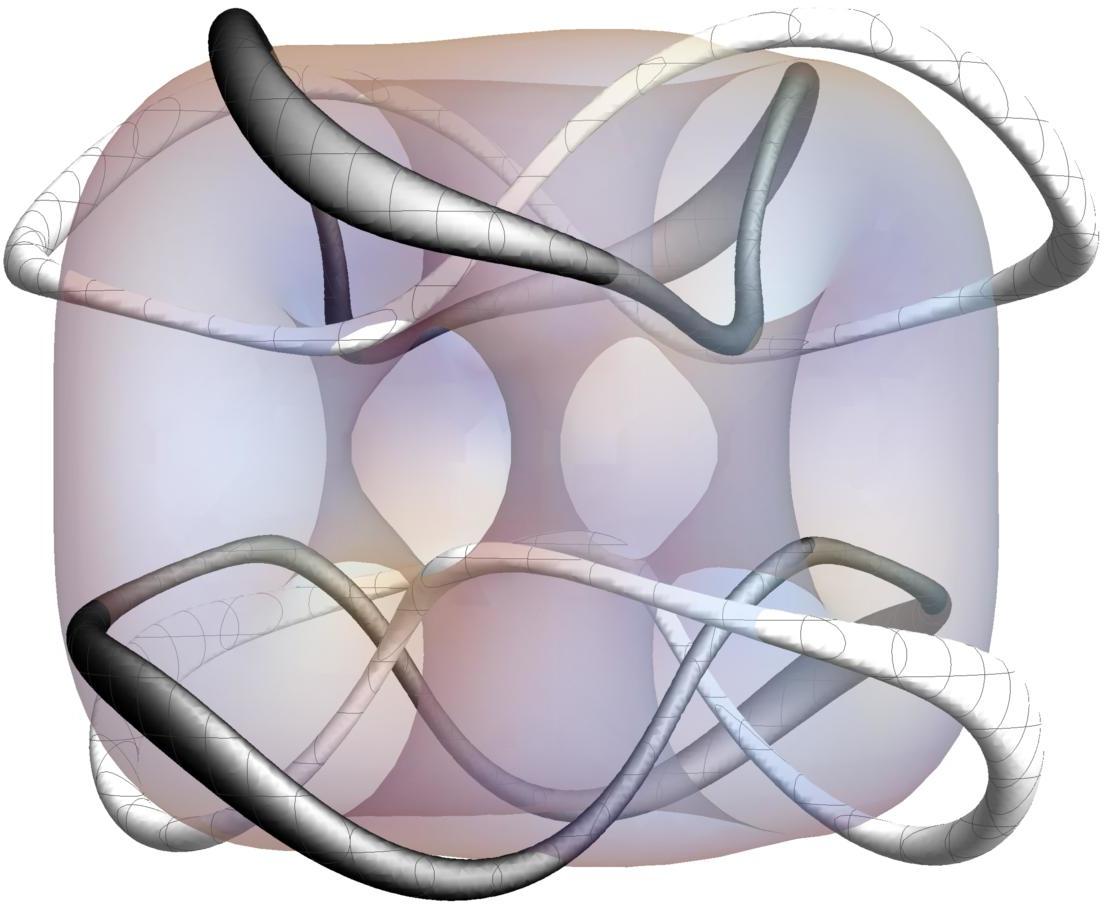}}}
\caption{Links for the $B=4$ Skyrmion.
(a) A link between the vortex ring (yellow) and the vacuum vortex
(magenta), which are both degenerate. (b,c) Nondegenerate links between the
vortices and vacuum vortices, which are both closed loops.
The gray isosurface is the baryon charge density illustrating the
shape of the Skyrmion.
}
\label{fig:RMB4}
\end{center}
\end{figure}

The $B=4$ Skyrmion has octahedral symmetry, which is the dual symmetry
of the cube, and the rational map with such symmetry
reads \cite{Houghton:1997kg}
\beq
R_4(z) = \frac{z^4 + \i2\sqrt{3}z^2 + 1}{z^4 - \i2\sqrt{3}z^2 + 1}.
\eeq
Fig.~\ref{fig:RMB4} shows preimages of $\bphi_{1,2}$ as well as two
rotations thereof, by $\beta=\frac\pi6$ and by
$\beta=\frac{3\pi}{2}$.
As for all $B>1$, the vacuum vortex (magenta) in fig.~\ref{fig:RMB4}
is degenerate, but this time also the vortex or antivacuum (yellow) is
degenerate with merging points of the curves at each face of the
cube.
Rotating the vortex points by $\beta=\frac\pi6$ and by
$\beta=\frac{3\pi}{2}$, see fig.~\ref{fig:RMB4}(b,c), yields regular
points on the 2-sphere under the mapping and the links are clear.
This time, however, the linking number is split into two disjoint
clusters of links and the total linking number is given by $q_1=1$,
$p_1=2$, $q_2=1$, $p_2=2$ and hence $B=Q=\sum_{\ell=1}^2p_\ell q_\ell=4$,
as promised.
This example confirms conjecture \ref{cjt:1} and this time with 2
clusters adding up to the total linking number.

An interesting note is that one can see the structure of the $B=4$
cubic Skyrmion being composed by two tori, with one of them flipped
with respect to the other.

\begin{figure}[!htp]
\begin{center}
\mbox{\subfloat[$\bphi_{1,2}$]{\includegraphics[scale=0.105]{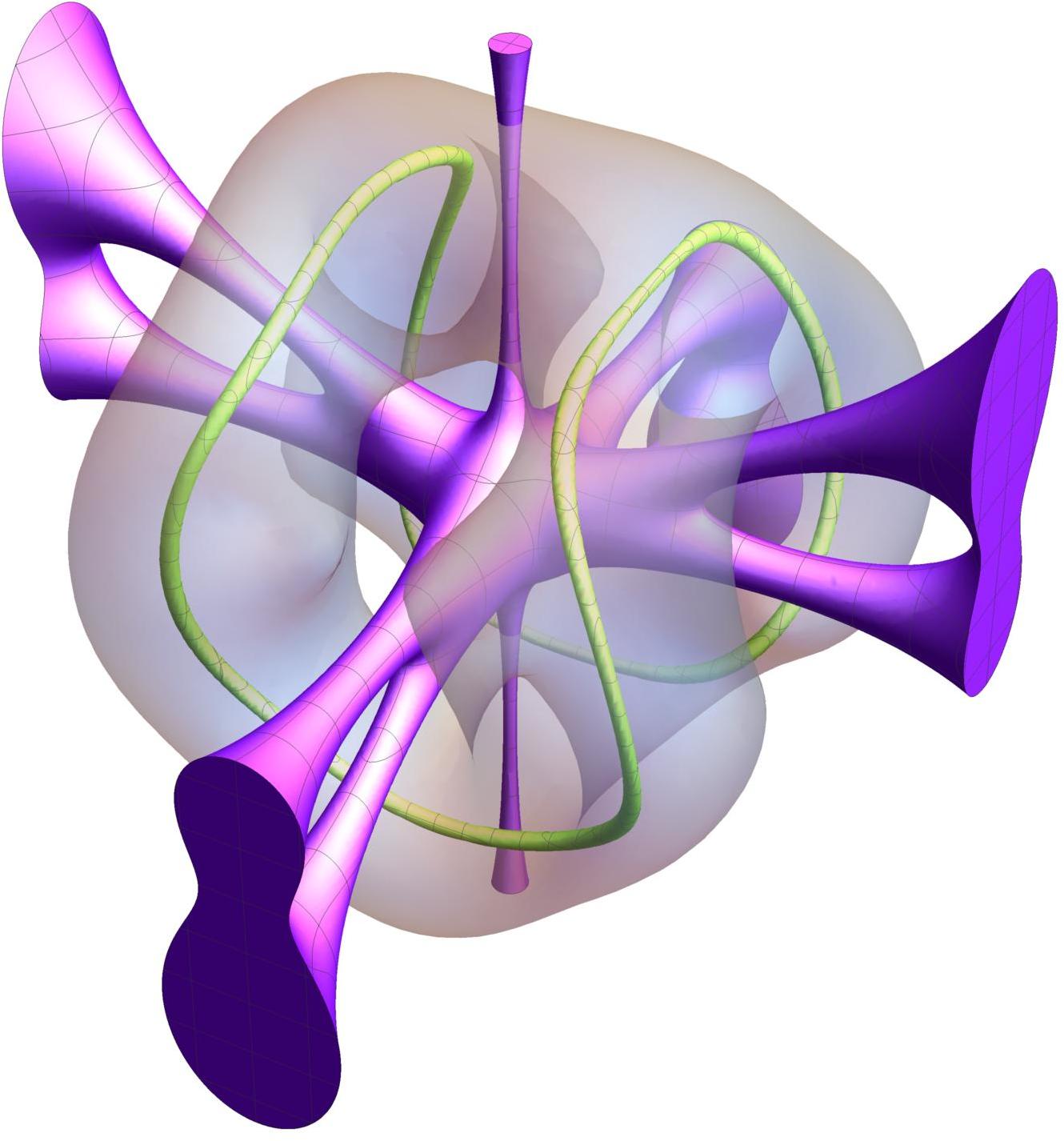}}
\subfloat[$\bphi_{1,2}^{M_{0\frac\pi20}}$]{\includegraphics[scale=0.12]{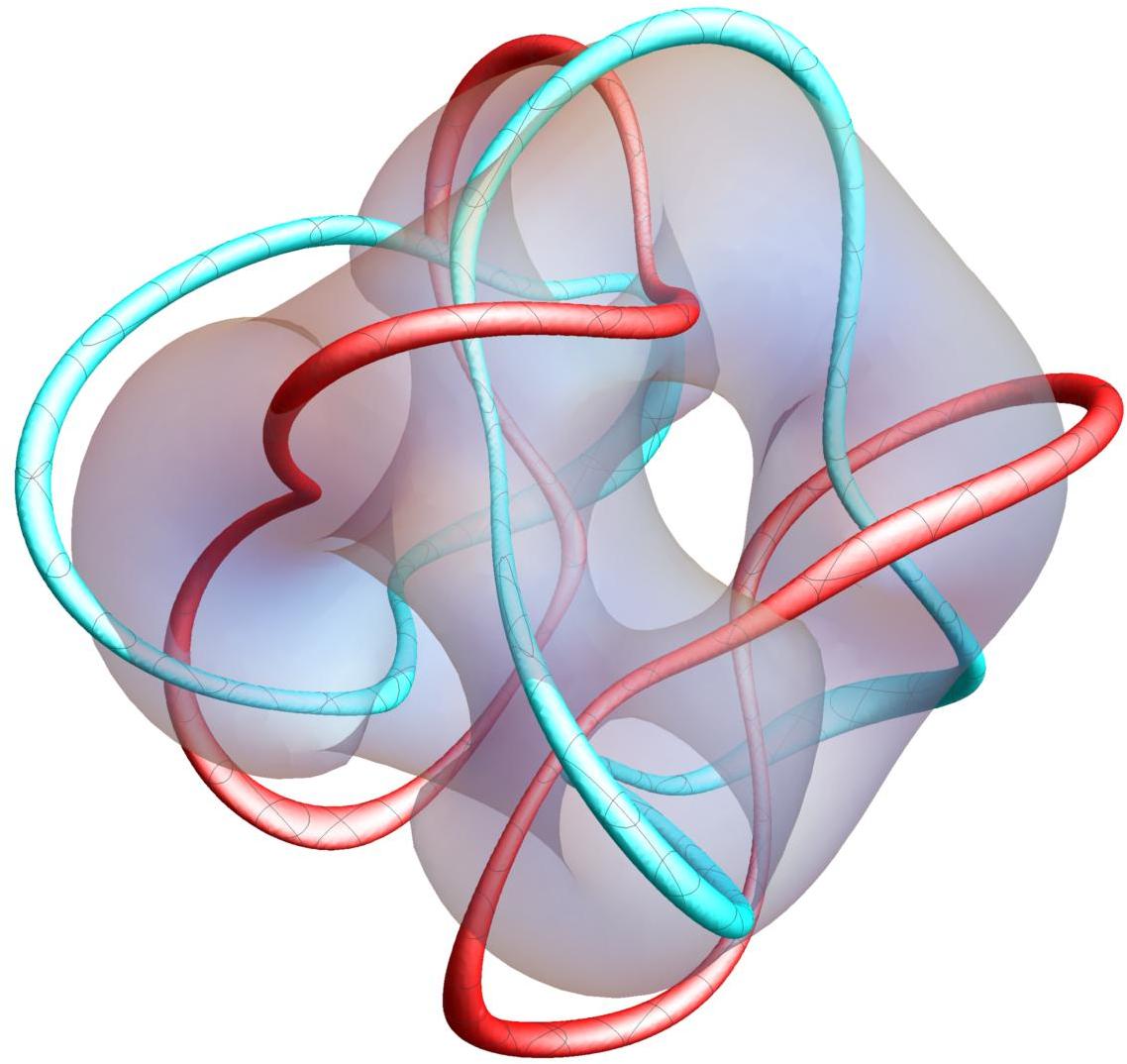}}
\subfloat[$\bphi_{1,2}^{M_{0\frac\pi2\frac\pi4}}$]{\includegraphics[scale=0.105]{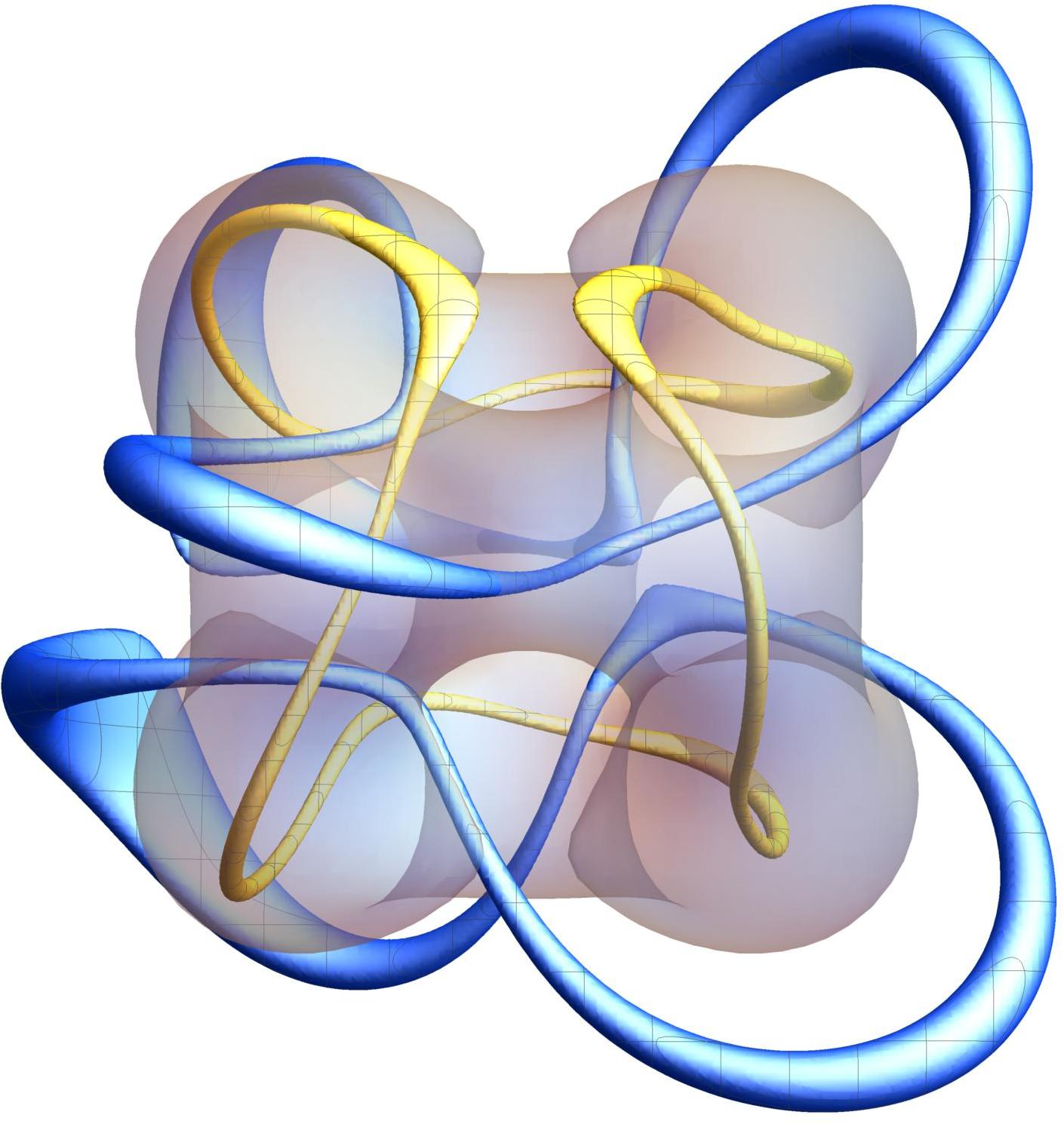}}}
\caption{Links for the $B=5$ Skyrmion.
(a) A link between the vortex ring (yellow) and the vacuum vortex
(magenta) which is degenerate. (b,c) Nondegenerate links between the
vortices and vacuum vortices, which are both closed loops.
The gray isosurface is the baryon charge density illustrating the
shape of the Skyrmion.
}
\label{fig:RMB5}
\end{center}
\end{figure}

The $B=5$ Skyrmion has dihedral ($D_{2d}$) symmetry and the
corresponding rational map is \cite{Houghton:1997kg}
\beq
R_5(z) = \frac{z\left(z^4 + b z^2 + a\right)}{a z^4 - b z^2 + 1},
\eeq
which contains enhanced $D_4$ symmetry if $b=0$ and further
enhancement to octahedral ($O_h$) symmetry if $a=-5$, see
ref.~\cite{Houghton:1997kg}.
The choice of the parameters is now for the first $B$ not fixed by
choosing the highest symmetry, because there is a lower value of
$\mathcal{I}$ (eq.~\eqref{eq:calI}) 
for different values of $a,b$.
In particular, $a=3.07$ and $b=3.94$ minimizes
$\mathcal{I}$ \cite{Houghton:1997kg}.

Fig.~\ref{fig:RMB5} shows preimages of $\bphi_{1,2}$ as well as two
rotations thereof by $\beta=\frac\pi2$ and by $\gamma=\frac\pi4$.
Only the vacuum vortex (magenta) is degenerate in the canonical frame,
see fig.~\ref{fig:RMB5}(a).
The easiest linking number is found in fig.~\ref{fig:RMB5}(c), where
the vortex (yellow) is linked twice with a vacuum vortex (blue) (bottom of
the figure) and thrice with another vacuum vortex (blue) (top of the
figure).
This yields $q=1$, $p=5$, yielding $B=Q=pq=5$, as expected.
The reason for counting five windings for the vacuum vortex is that
there is (from the vortex point of view) no difference between a
doubly wound vacuum vortex and two separate singly wound vacuum
vortices linking the vortex.
Hence, from the vortex point of view, there is a winding-5 vacuum
vortex that has split into two clusters (which is irrelevant for the
counting).
Of course, we would have taken the opposite point of view, reversing
the roles of the two preimages.
This would lead to $q_1=2$, $p_1=1$, $q_2=3$, $p_2=1$ and now
$B=Q=\sum_{\ell=1}^2p_\ell q_\ell=5$.

Turning to the counting of the linking number in
fig.~\ref{fig:RMB5}(b), the situation is slightly complicated by the
fact that the two clusters are linked.
Taking the viewpoint of the red vortices, we have $q_1=1$, $p_1=2$,
$q_2=1$, $p_2=3$ and $B=Q=5$ as promised.
If we swap the roles of the two preimages, we of course get the same
answer.

This is the first nontrivial example in the class of rational map
Skyrmions and it still confirms conjecture \ref{cjt:1}.

\begin{figure}[!htp]
\begin{center}
\mbox{\subfloat[$\bphi_{1,2}$]{\includegraphics[scale=0.105]{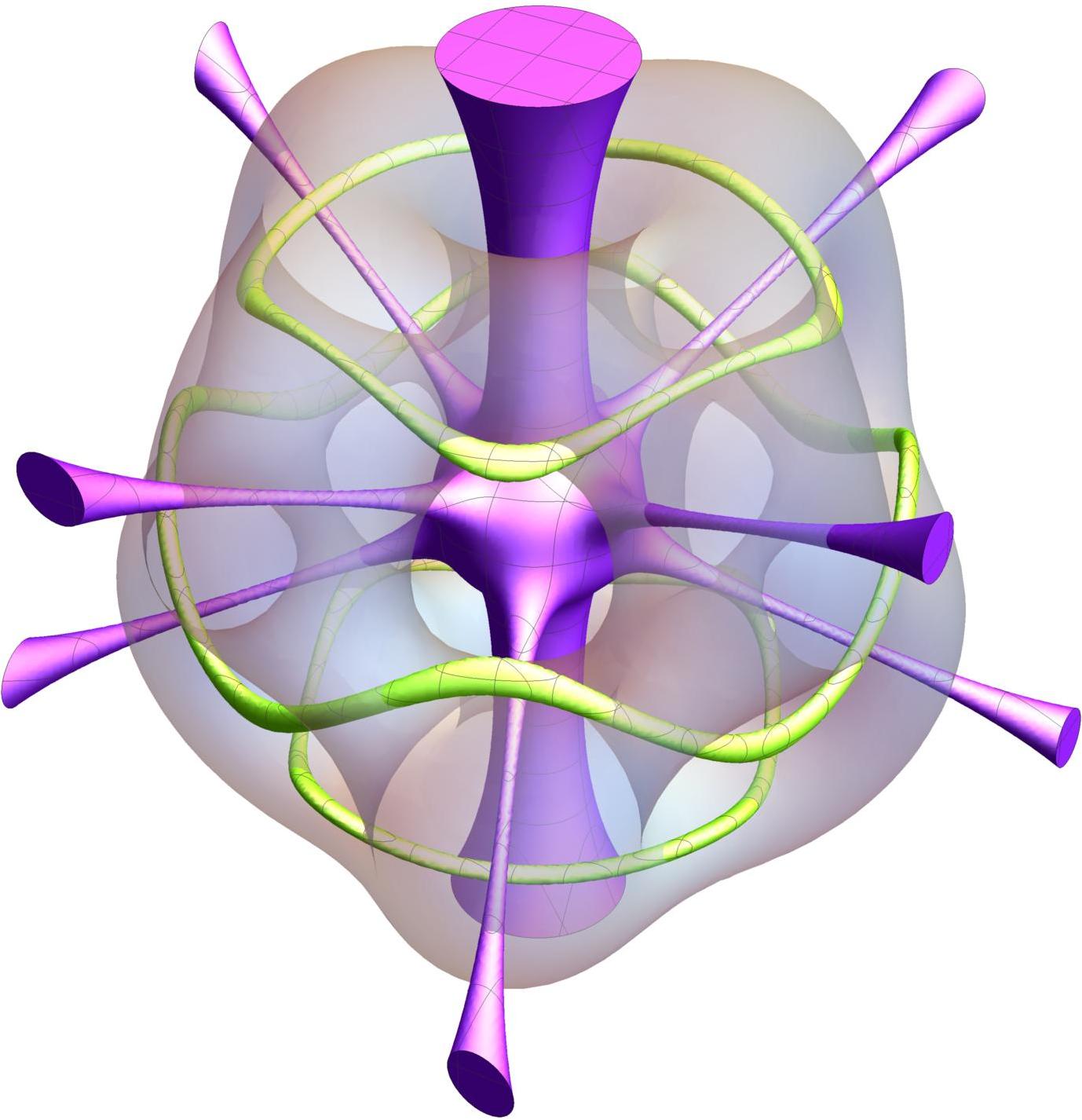}}
\subfloat[$\bphi_{1,2}^{M_{0\frac{3\pi}{2}0}}$]{\includegraphics[scale=0.145]{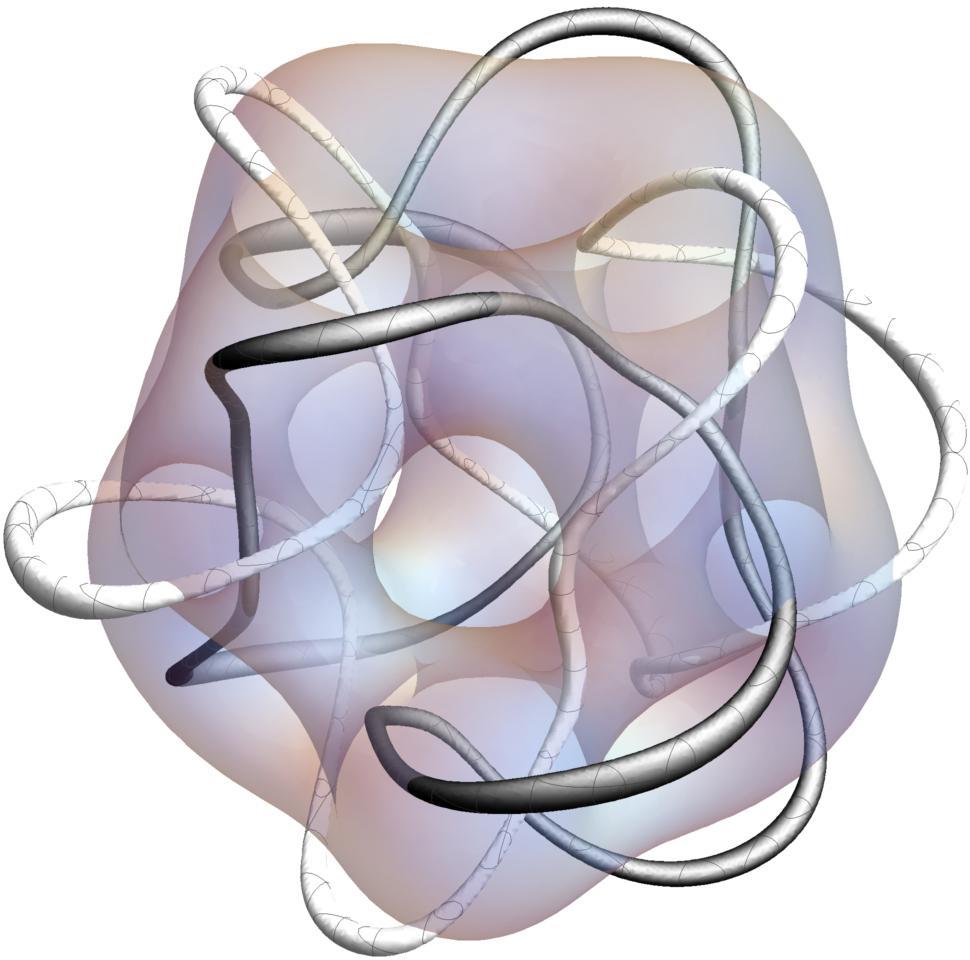}}}
\caption{Links for the $B=6$ Skyrmion.
(a) Links between the vortex rings (yellow) and the vacuum vortex
(magenta) which is degenerate. (b) Nondegenerate links between the
vortex and vacuum vortex, which are both closed loops.
The gray isosurface is the baryon charge density illustrating the
shape of the Skyrmion.
}
\label{fig:RMB6}
\end{center}
\end{figure}

The $B=6$ Skyrmion has $D_{4d}$ dihedral symmetry, which is generated
by the rational map \cite{Houghton:1997kg}
\beq
R_6(z) = \frac{z^4 + \i a}{z^2(\i a z^4 + 1)},
\eeq
with $a\in\mathbb{R}$.
This is the first $B$ for which symmetry does not fix the parameters
of the rational map.
Minimization of $\mathcal{I}$ (eq.~\eqref{eq:calI}) yields
$a=0.16$ \cite{Houghton:1997kg}. 

Fig.~\ref{fig:RMB6} shows preimages of $\bphi_{1,2}$ as well as of
a rotation of them by $\beta=\frac{3\pi}{2}$.
The vacuum vortex (magenta) in fig.~\ref{fig:RMB6}(a) is still
degenerate as promised, but after a swift $\beta$ rotation, the
mapping of the vortex points is regular.
After a bit of disentangling, it is clear that the vacuum vortex
(white) links the vortex (black) six times in fig.~\ref{fig:RMB6}(b),
corresponding to $q=1$, $p=6$ and $B=Q=pq=6$.

\begin{figure}[!htp]
\begin{center}
\mbox{\subfloat[$\bphi_{1,2}$]{\includegraphics[scale=0.093]{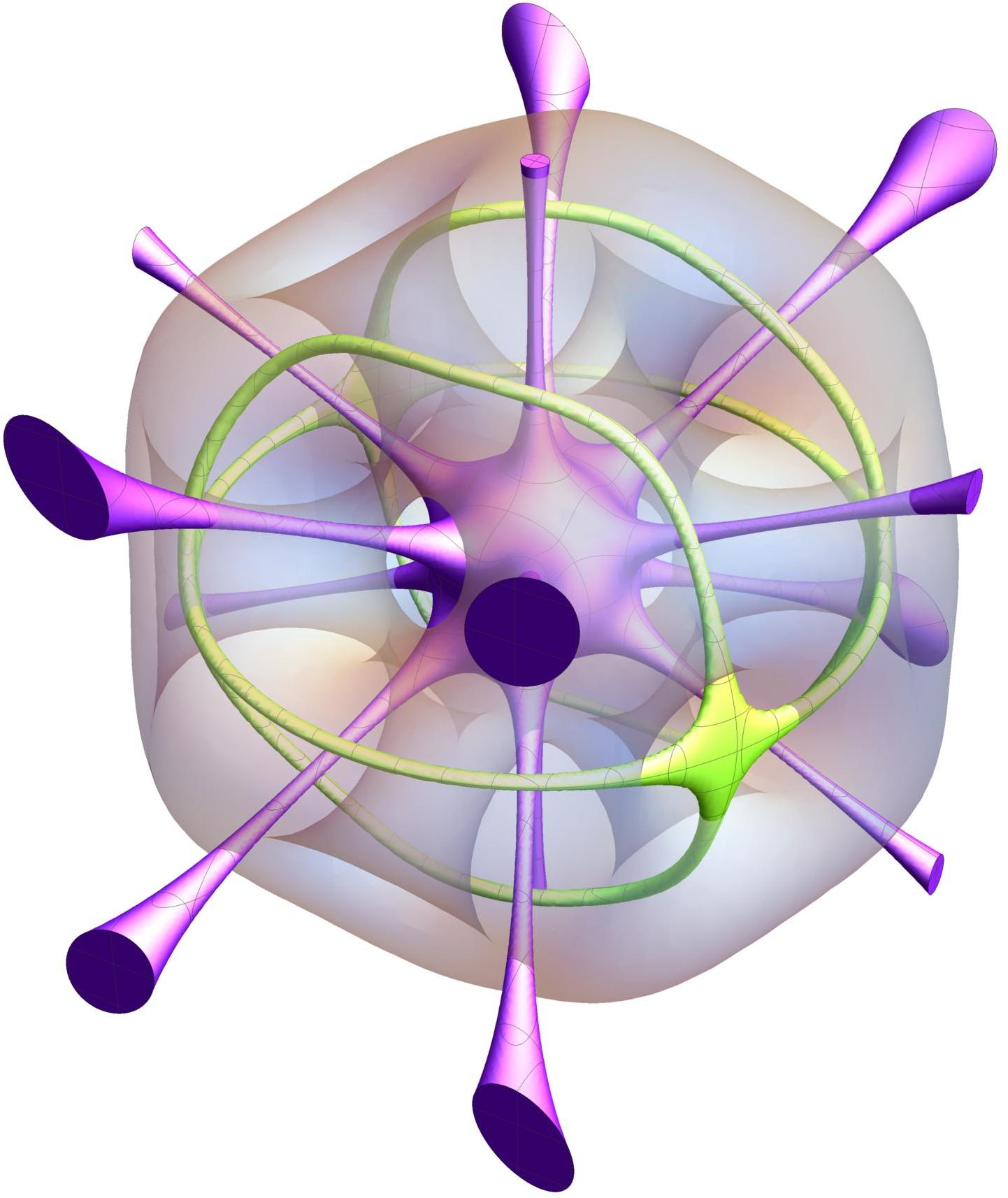}}
\subfloat[$\bphi_{1,2}^{M_{0\frac\pi60}}$]{\includegraphics[scale=0.153]{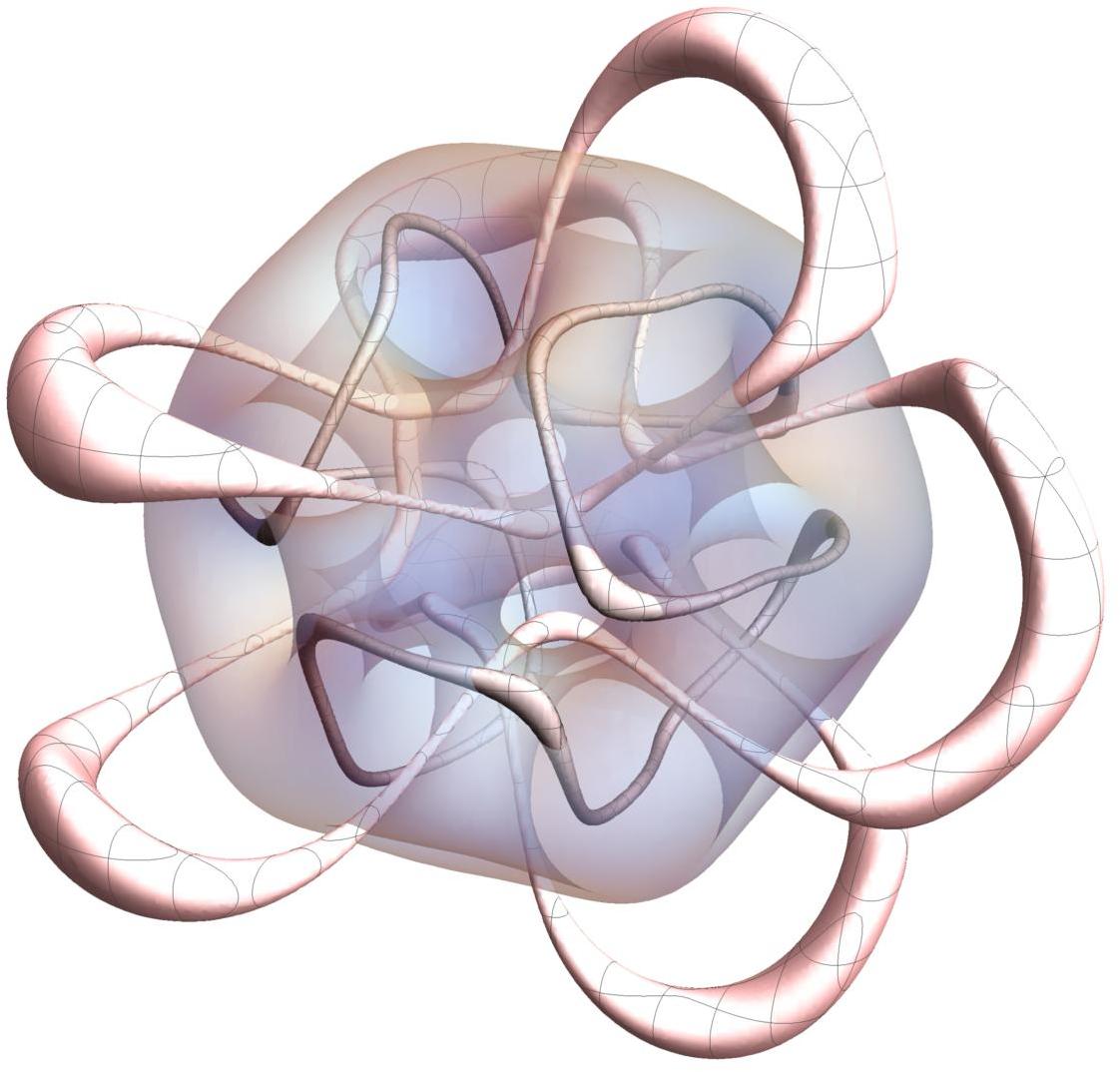}}
\subfloat[$\bphi_{1,2}^{M_{0\frac\pi20}}$]{\includegraphics[scale=0.108]{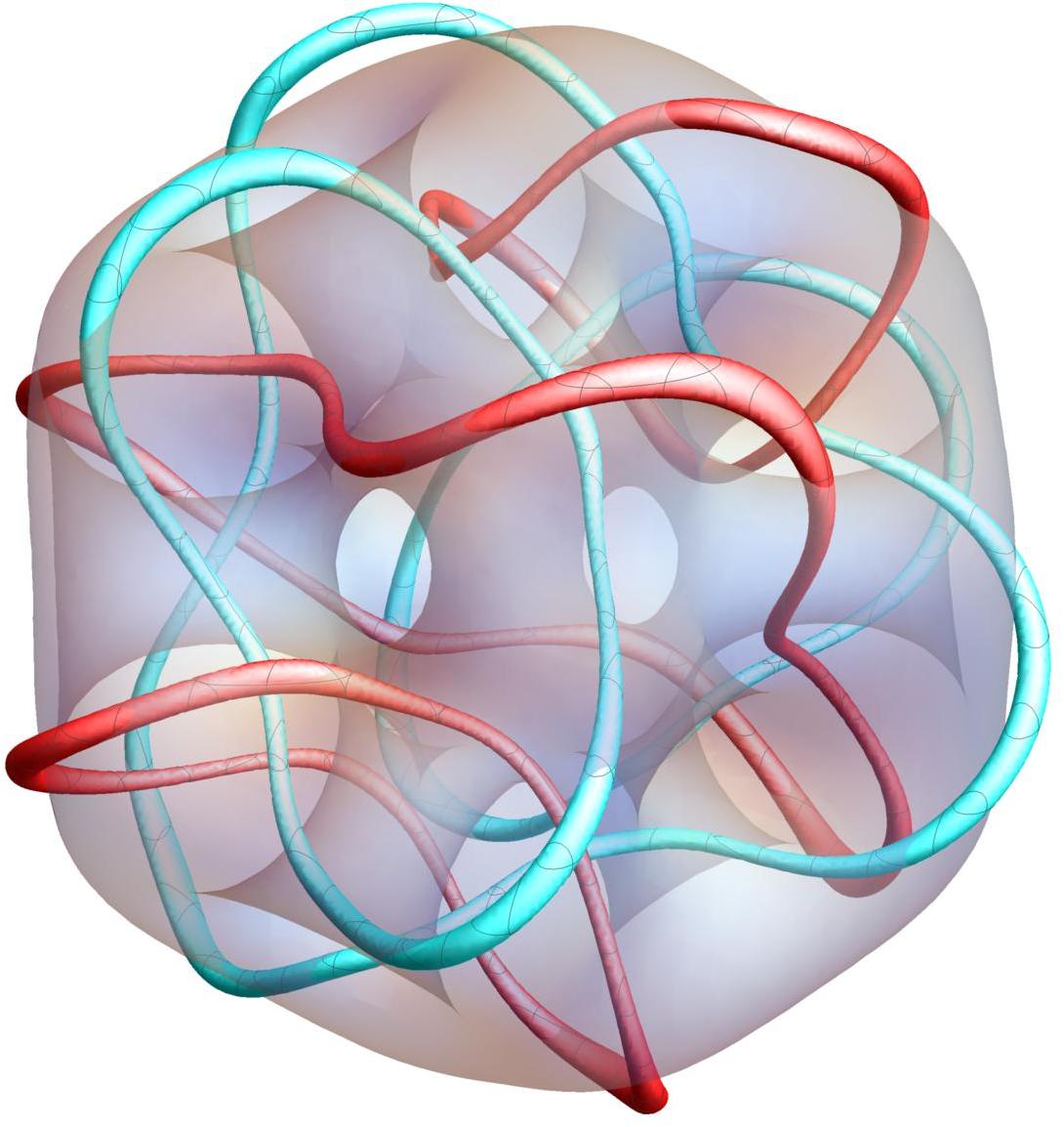}}}
\caption{Links for the $B=7$ Skyrmion.
(a) A link between the vortex ring (yellow) and the vacuum vortex
(magenta), which are both degenerate. (b,c) Nondegenerate links between the
vortices and vacuum vortices, which are both closed loops.
The gray isosurface is the baryon charge density illustrating the
shape of the Skyrmion.
}
\label{fig:RMB7}
\end{center}
\end{figure}

The $B=7$ Skyrmion is the most symmetric of them all and possesses
icosahedral symmetry, which fixes the rational map
as \cite{Houghton:1997kg} 
\beq
R_7(z) = \frac{z^5 + 3}{z^2(3z^5 + 1)}.
\eeq
Fig.~\ref{fig:RMB7} shows preimages of $\bphi_{1,2}$ as well as of
rotations thereof by $\beta=\frac\pi6$ and by $\beta=\frac\pi2$. 
In the canonical frame, both vortices are degenerate.
After rotating by $\beta=\frac\pi6$ the mapping is regular and the
vortex (dark red) links three vacuum vortices (ligth red) two, three
and two times, respectively, yielding $q=1$, $p=7$, $B=Q=pq=7$.
The counting goes slightly different if we continue the rotation of
the 2-sphere to $\beta=\frac\pi2$ where both vortices have turned into
3 rings.
If we take the point of view of the red vortices, the winding numbers
are $q_1=1$, $p_1=2$, $q_2=1$, $p_2=3$, $q_3=1$, $p_3=2$.
Notice, however, that the clusters themselves are linked and therefore
the number of vacuum vortices is not 7 but 3.

\begin{figure}[!htp]
\begin{center}
\mbox{\subfloat[$\bphi_{1,2}$]{\includegraphics[scale=0.095]{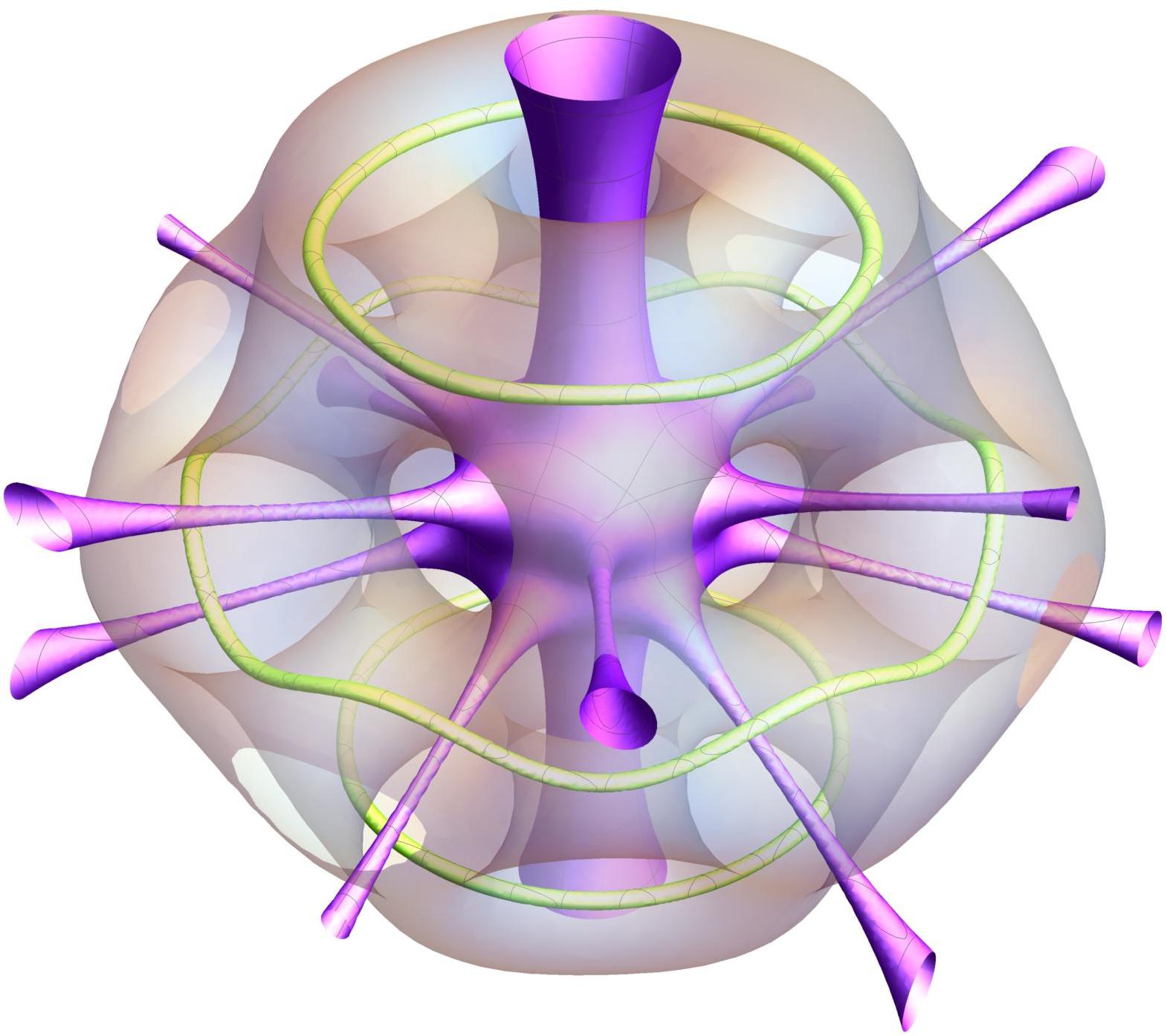}}
\subfloat[$\bphi_{1,2}^{M_{0\frac\pi20}}$]{\includegraphics[scale=0.125]{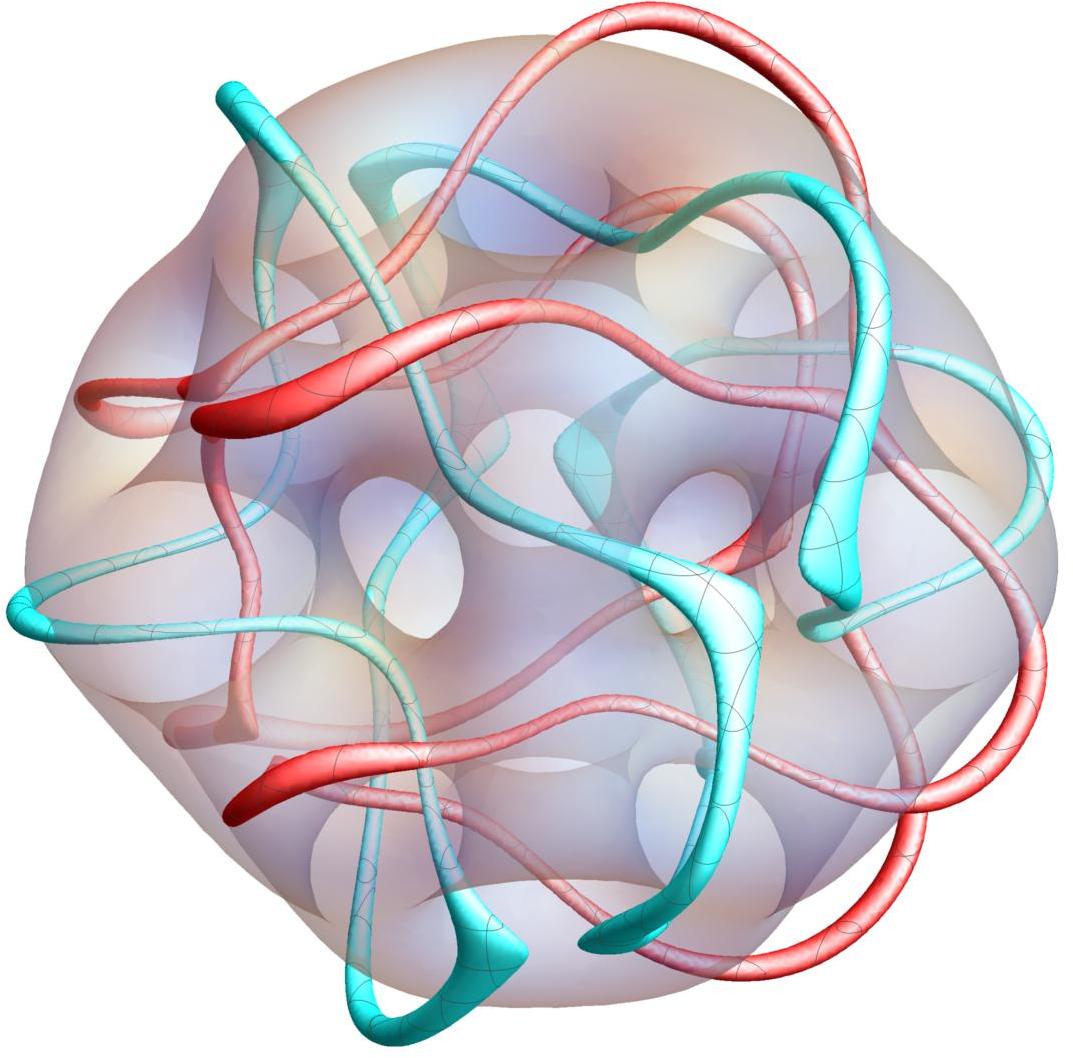}}}
\mbox{\subfloat[$\bphi_{1,2}^{M_{0\frac\pi2\frac\pi4}}$]{\includegraphics[scale=0.13]{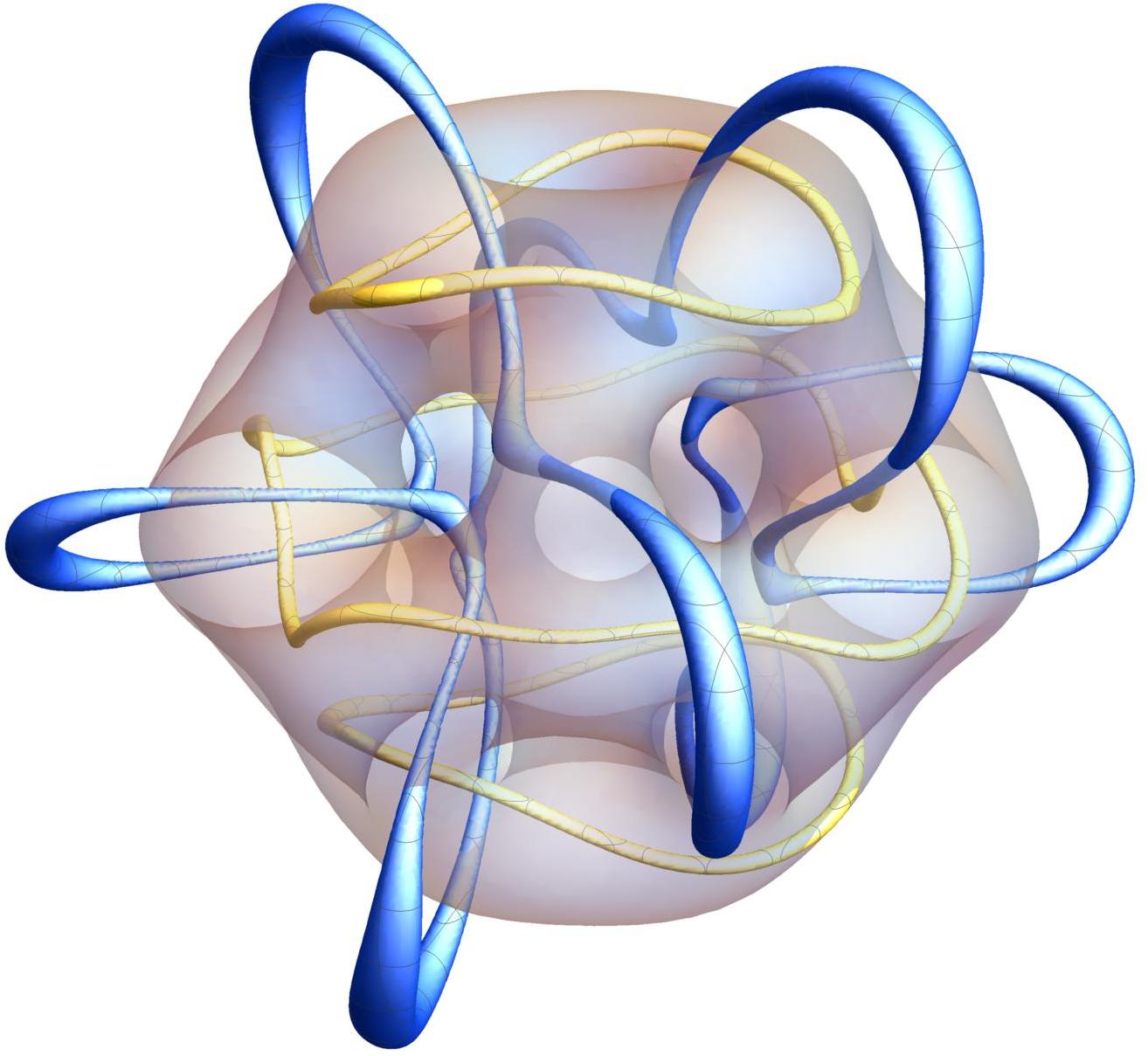}}}
\caption{Links for the $B=8$ Skyrmion.
(a) Links between the vortex rings (yellow) and the vacuum vortex
(magenta) which is degenerate. (b,c) Nondegenerate links between the
vortices and vacuum vortices, which are both closed loops.
The gray isosurface is the baryon charge density illustrating the
shape of the Skyrmion.
}
\label{fig:RMB8}
\end{center}
\end{figure}

The last Skyrmion here is the $B=8$ Skyrmion, which has $D_{6d}$
symmetry in the massless theory \eqref{eq:massless_Skyrme}, in
contradistinction from the solution of the massive theory which is
composed by two cubes \cite{Battye:2006na}.
The rational map for the fullerene-like Skyrmion with $D_{6d}$
symmetry has the corresponding rational map \cite{Houghton:1997kg}
\beq
R_8(z) = \frac{z^6 - a}{z^2(a z^6 + 1)},
\eeq
with $a\in\mathbb{R}$.
The minimization of $\mathcal{I}$ of eq.~\eqref{eq:calI} yields
$a=0.14$ \cite{Houghton:1997kg}. 

Fig.~\ref{fig:RMB8} shows preimages of $\bphi_{1,2}$ as well as of
rotations thereof by $\beta=\frac\pi2$ and by $\gamma=\frac\pi4$. 
As expected by now, the vacuum vortex in fig.~\ref{fig:RMB8}(a) is
degenerate.
In fig.~\ref{fig:RMB8}(b) the vacuum vortex (cyan) and the vortex
(red) are linked eight times and the counting is simply $q=1$, $p=8$,
and thus $B=Q=pq=8$.
Rotating by $\pi/4$ around the equator of the 2-sphere, yields
different preimages.
Now there are three vortices (yellow), see fig.~\ref{fig:RMB8}(c),
that link the vacuum vortex (blue) and they link the vacuum vortex
two, four and two times, respectively. 
The counting now goes like $q_1=1$, $p_1=2$, $q_2=1$, $p_2=4$,
$q_3=1$, $p_3=2$ and thus we have $B=Q=\sum_{\ell=1}^3 p_\ell q_\ell=8$
again, as promised.

\section{Discussion and outlook}\label{sec:discussion}

In this paper, we have proved theorem \ref{thm:1} which states that
the degree of a Skyrme field is the same as the linking number of two
preimages of two distinct regular points on the 2-sphere of said field
under the Hopf map.
We further conjecture that the 2 linked lines may be interpreted as
vortices in the original O(4) field.
Note that such an interpretation is impossible in the Faddeev-Skyrme
model which is based on O(3) fields in $\mathbb{R}^3$, although they
do possess Hopf charge and knots.

We illustrated the conjecture and hence the theorem with two examples:
a toroidal vortex, which is simply an axially symmetric Skyrmion with
topological degree $P$ (energetically stabilized by a certain
potential, see ref.~\cite{Gudnason:2016yix}); and with the eight first
rational map Skyrmions of ref.~\cite{Houghton:1997kg}.

The toroidal vortex or the $P$-wound axially symmetric Skyrmion is in
fact the motivation for conjecture \ref{cjt:1} and naturally it works
well.
The rational map Skyrmions, on the other hand, are a nontrivial check
on the conjecture and so far it has passed the checks.

One should note that all the preimages that we studied in this paper
are themselves unknots, viz.~they are topologically equivalent to
circles.
So we have only checked the conjecture \ref{cjt:1} with various
numbers of linked unknots.
It is possible that the conjecture needs refinement in more
complicated situations where the preimage itself become links or
a knot or even linked knots, which then by the nature of the game will
be linked with the other preimage.
Although $B=Q$ holds by theorem \ref{thm:1}, the conjecture may
receive corrections of the form, schematically
\begin{equation}
B = Q = \sum_{\rm unknots} p q
+ \sum_{\rm links} F_{\rm links}(p,q)
+ \sum_{\rm knots} F_{\rm knots}(p,q)
+ \sum_{\rm linked\ knots} F_{\rm linked\ knots}(p,q),
\end{equation}
where we have suppressed cluster indices.
We leave this for future studies.

Lord Kelvin imagined that atoms are described by
knots of vortices \cite{Thomson:1869}. 
With theorem \ref{thm:1} we can say that nuclei are not knots, but
contain links of vortices via a certain projection.

\subsection*{Acknowledgments}

We would like to thank Michikazu Kobayashi for collaboration at the
early stage of this work and we thank Chris Halcrow, Steffen Krusch,
Martin Speight and Paul Sutcliffe for discussions and comments. 
S.~B.~G. thanks the Outstanding Talent Program of Henan University for
partial support.
The work of S.~B.~G.~is supported by the National Natural Science
Foundation of China (Grant No.~11675223). 
M.~N.~is supported by the Ministry of Education, Culture, Sports,
Science (MEXT)-Supported Program for the Strategic Research Foundation
at Private Universities ``Topological Science'' (Grant No.~S1511006) and
by a Grant-in-Aid for Scientific Research on Innovative Areas
``Topological Materials Science'' (KAKENHI Grant No.~15H05855) from
MEXT, Japan.
M.~N.~is also supported in part by the Japan Society for
the Promotion of Science (JSPS) Grant-in-Aid for Scientific Research
(KAKENHI Grant No.~16H03984 and No.~18H01217).


\begin{thebibliography}{99}

\bibitem{Manton:2004}
  N.~Manton and P.~Sutcliffe,
  ``Topological Solitons,''
  Cambridge University Press (2004).

\bibitem{Ward:2001vi} 
  R.~S.~Ward,
  ``Hopf solitons from instanton holonomy,''
  \href{http://dx.doi.org/10.1088/0951-7715/14/6/307}{Nonlinearity {\bf 14}, 1543 (2001)}
  [\href{http://www.arxiv.org/abs/hep-th/0108082}{hep-th/0108082}].

\bibitem{Battye:1998pe} 
  R.~A.~Battye and P.~M.~Sutcliffe,
  ``Knots as stable soliton solutions in a three-dimensional classical field theory.,''
  \href{http://dx.doi.org/10.1103/PhysRevLett.81.4798}{Phys.\ Rev.\ Lett.\  {\bf 81}, 4798 (1998)}
  [\href{http://www.arxiv.org/abs/hep-th/9808129}{hep-th/9808129}].

\bibitem{Battye:1998zn} 
  R.~A.~Battye and P.~Sutcliffe,
  ``Solitons, links and knots,''
  \href{http://dx.doi.org/10.1098/rspa.1999.0502}{Proc.\ Roy.\ Soc.\ Lond.\ A {\bf 455}, 4305 (1999)}
  [\href{http://www.arxiv.org/abs/hep-th/9811077}{hep-th/9811077}].

\bibitem{Faddeev:1996zj} 
  L.~D.~Faddeev and A.~J.~Niemi,
  ``Knots and particles,''
  \href{http://dx.doi.org/10.1038/387058a0}{Nature {\bf 387}, 58 (1997)}
  [\href{http://www.arxiv.org/abs/hep-th/9610193}{hep-th/9610193}].

\bibitem{Meissner:1985nb} 
  U.~G.~Meissner,
  ``Toroidal Solitons With Unit Hopf Charge,''
  \href{http://dx.doi.org/10.1016/0370-2693(85)90582-9}{Phys.\ Lett.\  {\bf 154B}, 190 (1985).}

\bibitem{Ward:2004gr} 
  R.~S.~Ward,
  ``Skyrmions and Faddeev-Hopf solitons,''
  \href{http://dx.doi.org/10.1103/PhysRevD.70.061701}{Phys.\ Rev.\ D {\bf 70}, 061701 (2004)}
  [\href{http://www.arxiv.org/abs/hep-th/0407245}{hep-th/0407245}].

\bibitem{Gladikowski:1996mb} 
  J.~Gladikowski and M.~Hellmund,
  ``Static solitons with nonzero Hopf number,''
  \href{http://dx.doi.org/10.1103/PhysRevD.56.5194}{Phys.\ Rev.\ D {\bf 56}, 5194 (1997)}
  [\href{http://www.arxiv.org/abs/hep-th/9609035}{hep-th/9609035}].

\bibitem{Gudnason:2016yix} 
  S.~B.~Gudnason and M.~Nitta,
  ``Skyrmions confined as beads on a vortex ring,''
  \href{http://dx.doi.org/10.1103/PhysRevD.94.025008}{Phys.\ Rev.\ D {\bf 94}, no. 2, 025008 (2016)}
  [\href{http://www.arxiv.org/abs/arXiv:1606.00336}{arXiv:1606.00336 [hep-th]}].

\bibitem{Gudnason:2014gla} 
  S.~B.~Gudnason and M.~Nitta,
  ``Effective field theories on solitons of generic shapes,''
  \href{http://dx.doi.org/10.1016/j.physletb.2015.05.062}{Phys.\ Lett.\ B {\bf 747}, 173 (2015)}
  [\href{http://www.arxiv.org/abs/arXiv:1407.2822}{arXiv:1407.2822 [hep-th]}].

\bibitem{Gudnason:2014hsa} 
  S.~B.~Gudnason and M.~Nitta,
  ``Incarnations of Skyrmions,''
  \href{http://dx.doi.org/10.1103/PhysRevD.90.085007}{Phys.\ Rev.\ D {\bf 90}, no. 8, 085007 (2014)}
  [\href{http://www.arxiv.org/abs/arXiv:1407.7210}{arXiv:1407.7210 [hep-th]}].
  
\bibitem{Gudnason:2014jga} 
  S.~B.~Gudnason and M.~Nitta,
  ``Baryonic torii: Toroidal baryons in a generalized Skyrme model,''
  \href{http://dx.doi.org/10.1103/PhysRevD.91.045027}{Phys.\ Rev.\ D {\bf 91}, no. 4, 045027 (2015)}
  [\href{http://www.arxiv.org/abs/arXiv:1410.8407}{arXiv:1410.8407 [hep-th]}].

\bibitem{Gudnason:2018oyx} 
  S.~B.~Gudnason and M.~Nitta,
  ``Baryonic handles: Skyrmions as open vortex strings on a domain wall,''
  \href{http://dx.doi.org/10.1103/PhysRevD.98.125002}{Phys.\ Rev.\ D {\bf 98}, no. 12, 125002 (2018)}
  [\href{http://www.arxiv.org/abs/arXiv:1809.01025}{arXiv:1809.01025 [hep-th]}].

\bibitem{Kasamatsu}
  K. Kasamatsu, M. Tsubota, and M. Ueda,
  ``Vortices in multicomponent Bose–Einstein condensates,''
  \href{http://dx.doi.org/10.1142/S0217979205029602}{Int.\ J.\ Mod.\ Phys.\ B{\bf 19}, 1835 (2005)}
  [\href{http://www.arxiv.org/abs/cond-mat/0505546}{cond-mat/0505546}].

\bibitem{Ruostekoski:2001fc} 
  J.~Ruostekoski and J.~R.~Anglin,
  ``Creating vortex rings and three-dimensional skyrmions in Bose-Einstein condensates,''
  \href{http://dx.doi.org/10.1103/PhysRevLett.86.3934}{Phys.\ Rev.\ Lett.\  {\bf 86}, 3934 (2001)}
  [\href{http://www.arxiv.org/abs/cond-mat/0103310}{cond-mat/0103310}].

\bibitem{Battye:2001ec} 
  R.~A.~Battye, N.~R.~Cooper and P.~M.~Sutcliffe,
  ``Stable skyrmions in two component Bose-Einstein condensates,''
  \href{http://dx.doi.org/10.1103/PhysRevLett.88.080401}{Phys.\ Rev.\ Lett.\  {\bf 88}, 080401 (2002)}
  [\href{http://www.arxiv.org/abs/cond-mat/0109448}{cond-mat/0109448}].

\bibitem{Nitta:2012hy} 
  M.~Nitta, K.~Kasamatsu, M.~Tsubota and H.~Takeuchi,
  ``Creating vortons and three-dimensional skyrmions from domain wall annihilation with stretched vortices in Bose-Einstein condensates,''
  \href{http://dx.doi.org/10.1103/PhysRevA.85.053639}{Phys.\ Rev.\ A {\bf 85}, 053639 (2012)}
  [\href{http://www.arxiv.org/abs/arXiv:1203.4896}{arXiv:1203.4896 [cond-mat.quant-gas]}].

\bibitem{Houghton:1997kg} 
  C.~J.~Houghton, N.~S.~Manton and P.~M.~Sutcliffe,
  ``Rational maps, monopoles and Skyrmions,''
  \href{http://dx.doi.org/10.1016/S0550-3213(97)00619-6}{Nucl.\ Phys.\ B {\bf 510}, 507 (1998)}
  [\href{http://www.arxiv.org/abs/hep-th/9705151}{hep-th/9705151}].

\bibitem{Battye:2006na} 
  R.~Battye, N.~S.~Manton and P.~Sutcliffe,
  ``Skyrmions and the alpha-particle model of nuclei,''
  \href{http://dx.doi.org/10.1098/rspa.2006.1767}{Proc.\ Roy.\ Soc.\ Lond.\ A {\bf 463}, 261 (2007)}
  [\href{http://www.arxiv.org/abs/hep-th/0605284}{hep-th/0605284}].

\bibitem{Thomson:1869}
  W.~H.~Thomson,
  ``On Vortex Motion,''
  \href{https://doi.org/10.1017/S0080456800028179}{Trans.\ R.\ Soc.\ Edin.\ {\bf 25}, 217 (1868).}


\end{thebibliography}
\end{document}